\documentclass[fleqn,usenatbib]{mnras}

\usepackage{newtxtext,newtxmath}
 \usepackage{booktabs}
 \usepackage{graphicx}

\usepackage[T1]{fontenc}

\DeclareRobustCommand{\VAN}[3]{#2}
\let\VANthebibliography\thebibliography
\def\thebibliography{\DeclareRobustCommand{\VAN}[3]{##3}\VANthebibliography}

\usepackage{graphicx}	
\usepackage{amsmath}	
\usepackage{hyperref}
\usepackage{subcaption}
\hypersetup{unicode=true, colorlinks=true, linkcolor=[rgb]{0.53, 0.15, 0.34}, citecolor=[rgb]{0.0, 0.27, 0.42}, filecolor=[rgb]{1.0, 0.13, 0.32}, urlcolor=[rgb]{0.53, 0.15, 0.34}}

\usepackage{color}
\usepackage{array}
\usepackage[dvipsnames]{xcolor}

\title[Reflection and Photoionised Absorption in GX 13+1]{Relativistic X-Ray Reflection and Photoionised Absorption in the Neutron-Star Low-Mass X-ray Binary GX 13+1}

\author[Saavedra et al.]{
Enzo A. Saavedra$^{1}$\thanks{e-mail: enzosaave@fcaglp.unlp.edu.ar},
Federico García$^{1,2}$,
Federico A. Fogantini$^{2}$,
Mariano Méndez$^{3}$,
\and
Jorge A. Combi$^{1,2,4}$,
Pedro L. Luque-Escamilla$^{5}$, 
Josep Martí$^{4}$
\\
$^{1}$Facultad de Ciencias Astron\'omicas y Geof\'{\i}sicas, Universidad Nacional de La Plata, Paseo del Bosque, B1900FWA La Plata, Argentina\\
$^{2}$Instituto Argentino de Radioastronom\'ia (CCT La Plata, CONICET; CICPBA; UNLP), C.C.5, (1894) Villa Elisa, Buenos Aires, Argentina\\
$^{3}$Kapteyn Astronomical Institute, University of Groningen, PO BOX 800, NL-9700 AV Groningen, the Netherlands \\
$^{4}$Departamento de F\'isica (EPS), Universidad de Ja\'en, Campus Las Lagunillas s/n, A3, 23071 Ja\'en, Spain \\
$^{5}$Departamento de Ingeniería Mecánica y Minera (EPSJ), Universidad de Jaén, Campus Las Lagunillas s/n Ed. A3, E-23071 Jaén,
Spain
}

\date{Accepted XXX. Received YYY; in original form ZZZ}

\pubyear{2015}

\begin{document}
\label{firstpage}
\pagerange{\pageref{firstpage}--\pageref{lastpage}}
\maketitle

\begin{abstract}
We analysed a dedicated {\it NuSTAR} observation of the neutron-star low-mass X-ray binary Z-source GX 13+1 to study the timing and spectral properties of the source. From the colour-colour diagram, we conclude that during that observation the source transitioned from the normal branch to the flaring branch. We fitted the spectra of the source in each branch with a model consisting of an accretion disc, a Comptonised blackbody, relativistic reflection ({\tt relxillNS}), and photo-ionised absorption ({\tt warmabs}). Thanks to the combination of the large effective area and good energy resolution of {\it NuSTAR} at high energies, we found evidence of relativistic reflection in both the Fe K line profile, and the Compton hump present in the 10--25~keV energy range. The inner disc radius is $R_{\rm in} \lesssim 9.6~r_g$, which allowed us to further constrain the magnetic field strength to  $B \lesssim 1.8 \times 10^8$~G.  We also found evidence for the presence of a hot wind leading to photo-ionised absorption of Fe and Ni, with a Ni overabundance of $\sim$6 times solar. From the spectral fits, we find that the distance between the ionising source and the slab of ionised absorbing material is $\sim 4-40 \times 10^5$~km. We also found that the width of the boundary layer extends $\sim$3~km above the surface of a neutron star, which yielded a neutron-star radius $R_{\rm NS}\lesssim 16$~km. The scenario inferred from the spectral modelling becomes self-consistent only for high electron densities in the accretion disk, $n_e \sim 10^{22}-10^{23}$~cm$^{-3}$, as expected for a Shakura-Sunyaev disc, and significantly above the densities provided by {\tt relxillNS} models.
These results have implications for our understanding of the physical conditions in GX 13+1.
\end{abstract}



\section{Introduction}

Neutron-stars in Low-Mass X-ray Binaries (NS-LMXBs) accrete matter from a low-mass companion star via Roche-lobe overflow \citep{tauris_vandenheuvel_2006, Zhu2012PASP..124..195Z}.
The spectra of NS-LMXBs are well described by a combination of a multi-colour blackbody component associated with the thermal emission from an accretion disk close to the NS and a high-energy Comptonised continuum that usually dominates the spectrum and is associated with the contribution of the NS surface and/or a corona. \citep{2005ApJS..157..335L, Koljonen2020A&A...639A..13K}. 
Part of these photons can be reprocessed by the accretion disk and re-emitted in the form of a continuum spectral component with a series of atomic features accompanied by a Compton back-scattering hump \citep{Rae2005MNRAS.364.1229R, Miller, Risaliti}, known as a reflection spectrum \citep{1988MNRAS.233..475G, 1988ApJ...331..939W, 1989MNRAS.238..729F, George1991MNRAS.249..352G}. 
The most prominent feature of this reflection is the Fe K emission line complex between 6.4--6.97~keV. The Fe line profile is shaped by Doppler and relativistic effects due to the rotational velocity of the disk and the strong gravitational field of the compact object \citep{Fabian1989MNRAS.238..729F, Miller2007ARA&A..45..441M}. 
The study of disk reflection spectra allows us to derive information of the physical and geometrical parameters of the system, such as the inner disk radius and the inclination of the accretion disk to the line of sight \citep{Dauser2010MNRAS.409.1534D, Miller2013ApJ...779L...2M, Degenaar2015MNRAS.451L..85D, Degenaar2016MNRAS.461.4049D, Mondal2019MNRAS.487.5441M, Anitra2021A&A...654A.160A}. The energy resolution and effective area provided by the {\it NuSTAR} observatory enable the detection of spectral features and Compton humps with high efficacy \citep{2013ApJ...770..103H}.

NS-LMXBs are divided into two classes: {\it Z} and {\it Atoll} sources \citep{1989A&A...225...79H, Muno2002ApJ...568L..35M}. This classification is based on the shape that the individual sources trace in the X-ray colour-colour diagram (CCD) or the hardness-intensity diagram (HID). The $Z$-sources form an approximate $Z$-path in the CCD and HID, showing three branches, so-called the horizontal branch (HB), the normal branch (NB), and the flaring branch \citep[FB; see, e.g.,][and references therein]{Shirey1999ApJ...517..472S, Lin2009ApJ...699...60L, Lin2012ApJ...756...34L, Ding2015JApA...36..335D, Mondal2018MNRAS.474.2064M, Homan2018ApJ...853..157H, Coughenour2018ApJ...867...64C, Mazzola2019A&A...621A..89M, Agrawal2020MNRAS.497.3726A}. The $Z$ and $\textit{Atoll}$ sources both have CCDs with similar shapes, consisting of three branches. The range of X-ray intensity and the time scale on which the sources move along the diagrams is significantly longer for $\textit{Atoll}$ sources than $Z$-sources, by one to two orders of magnitude \citep{Gierlinski2002MNRAS.331L..47G, Muno2002ApJ...568L..35M}. 

Two widely accepted scenarios exist for explaining the spectral properties of $\textit{Z}$-sources. The first scenario involves a model that comprises a multicolor blackbody emission from a standard accretion disc and a component arising from the inverse Compton scattering of soft seed photons by hot plasma in the boundary layer or central corona. \citep{Mitsuda1984PASJ...36..741M, Barret2001AdSpR..28..307B, Disalvo2002A&A...386..535D, Agrawal2003MNRAS.346..933A, Agrawal2009MNRAS.398.1352A}. Alternatively, another model suggests that the $\textit{Z}$-source spectrum can be explained by two components: a single blackbody representing the temperature of the boundary layer or surface of the neutron star and a hard Comptonised emission originating from the hot inner accretion flow \citep{DiSalvo2000ApJ...544L.119D, Disalvo2001ApJ...554...49D, Sleator2016ApJ...827..134S}. The key difference between these scenarios is the choice of the soft component, which can be either a multi-color disc emission or a blackbody emission from the surface of the neutron star. Despite using the same model, the Comptonised component may be associated with different regions of the accretion flow. While a combination of multi-color disc and blackbody emission can adequately explain the X-ray spectra of NS-LMXBs, this model is more appropriate for spectra lacking a strong hard tail or extended high energy coverage \citep{Lin2012ApJ...756...34L}.

In contrast to $\textit{Atoll}$ sources, which exhibit significant spectral changes \citep[see for e.g.][]{Schulz1989A&A...225...48S}, the spectral changes along the $\textit{Z}$-path are much more subtle. Although $\textit{Atoll}$ sources in the lower regions of their CCDs can sometimes display soft $\textit{Z}$-source-like energy spectra \citep[see for e.g.][]{Oosterbroek2001A&A...366..138O}, they can also exhibit a hard state with a relatively flat power-law energy spectrum with $\Gamma \sim 1.8$ in the range of 2$-$100~keV when they are weak \citep[][]{Barret2000ApJ...533..329B}. This type of hard spectrum is not observed in $\textit{Z}$-sources, possibly due to the fact that they are not observed at low luminosities.

The temporal properties of $\textit{Z}$-sources correlate with the position of the source in the CCD. These sources were observed to have three different types of quasi-periodic oscillations (QPOs). QPOs with a frequency of 15--100~Hz appear on the HB, and are therefore called HB oscillations (HBOs) \citep{1998ApJ...504L..35W, 2002ApJ...568..878H}. QPOs with a frequency of 5--15~Hz are observed both in the NB and the FB, and are called, respectively, NBO and FBOs \citep{2002ApJ...568..878H}. These low-frequency oscillations were observed in almost all Galactic $\textit{Z}$-sources, except for GX~349+2 \citep{2001ASPC..251..396O, 2003MNRAS.346..933A}. 
The third type of QPOs are the kiloHertz (kHz) QPOs, with frequencies ranging from 300 to 1200~Hz, which are observed in all branches. The frequency of kHz QPOs increases as the source moves along the $\textit{Z}$-path from the HB to the FB \citep{2000ARA&A..38..717V, Jackson2009A&A...494.1059J, Sanna2010MNRAS.408..622S, 2021ASSL..461..263M, Peirano2022MNRAS.513.2804P, Hiemstra2011MNRAS.411..137H}.

GX 13+1 is an NS-LMXB classified initially as a persistently bright $\textit{Atoll}$ source \citep{1989A&A...225...79H}, and later re-classified as a $\textit{Z}$-source due to strong secular evolution of its CCDs and HIDs \citep{Fridriksson}. Type-I X-ray bursts were observed for the first time in this system in 1985 using \textit{SAS}-3 \citep{Fleischman1985A&A...153..106F, Matsuba1995PASJ...47..575M}, which allowed the identification of the compact object as a NS. This system has an orbital period of $\sim$24.7~d \citep{2010ApJ...719..979C}, a companion star with a mass of $\sim$0.4~M$_\odot$, and is located at a distance of 7$\pm$1~kpc \citep{2002ApJ...570..793B}.

GX 13+1 also belongs to the class of dipping sources \citep{2014A&A...564A..62D}. These systems are believed to be observed close to the orbital plane, which causes periodic dips in the observed X-ray radiation caused by the obscuration of the central X-ray source by structures in the outer regions of the accretion disc. The characteristics of the dips, such as their depth, duration, and spectral evolution, can vary between different sources and even between different cycles of the same source.   \citet{Boirin2005A&A...436..195B} and \citet{DiazTrigo2006A&A...445..179D} showed that photo-ionised plasma plays a significant role in LMXBs. By modelling the changes in the narrow X-ray absorption features and the continuum observed by XMM-{\it Newton}, they demonstrated that these changes are due to a higher column density and a lower ionisation state of the absorbing plasma. The fact that dipping sources are normal LMXBs viewed from close to the orbital plane implies that photo-ionised plasma is a common feature of LMXBs. Outside of the dips, the properties of the absorbers do not vary significantly with the orbital phase, indicating that the ionised plasma has a cylindrical geometry with a maximum column density near the plane of the accretion disc. Highly ionised absorption lines such as Fe~XXV and Fe~XXVI trace winds, which are a fundamental component associated with the accretion process \citep[see][and references therein]{Ponti2015MNRAS.446.1536P}.

Several spectral studies of GX~13+1 were performed in the past using data from different satellites, such as {\it ASCA} \citep{2010ApJ...719..979C}, {\it RXTE} \citep{2004A&A...418..255H}, {\it Chandra} \citep{2005ApJ...620..274U} and INTEGRAL/{\it ISGRI} \citep{2010A&A...512A..57M}. \citet{2012A&A...543A..50D} analysed several XMM-{\it Newton} observations of this source and reported high absorption along the line of sight obscuring up to 80\% of the total emission. They concluded that the presence of a disk wind and/or a warm atmosphere may explain the observations of GX~13+1, leading to a system inclination of 60--80$^\circ$. In addition, \citet{Maiolino2019A&A...625A...8M} analysed the same XMM-{\it Newton} observations, fitting a {\tt diskline} type model to an Fe K$_{\alpha}$ asymmetric emission line, finding that the inclination was also $\sim$60$^\circ$.  

Here we report the first temporal and spectral X-ray analysis of GX~13+1 using broadband {\it NuSTAR} data. This paper is organised as follows: in \hyperref[sec:data]{Section~\ref{sec:data}} we describe the observational data and reduction procedures. In \hyperref[sec:timing]{Section~\ref{sec:timing}} we present the results of the timing analysis, while in \hyperref[sec:spectral]{Section~\ref{sec:spectral}} we present the results of spectral analyses. In \hyperref[sec:Discussion]{Section~\ref{sec:Discussion}} we discuss our results, and finally in \hyperref[sec:Conclusions]{Section~\ref{sec:Conclusions}} we summarise our main conclusions.

\section{Observations and Data Reduction} \label{sec:data}

 \begin{figure*}
    \centering
    \includegraphics[width=\columnwidth]{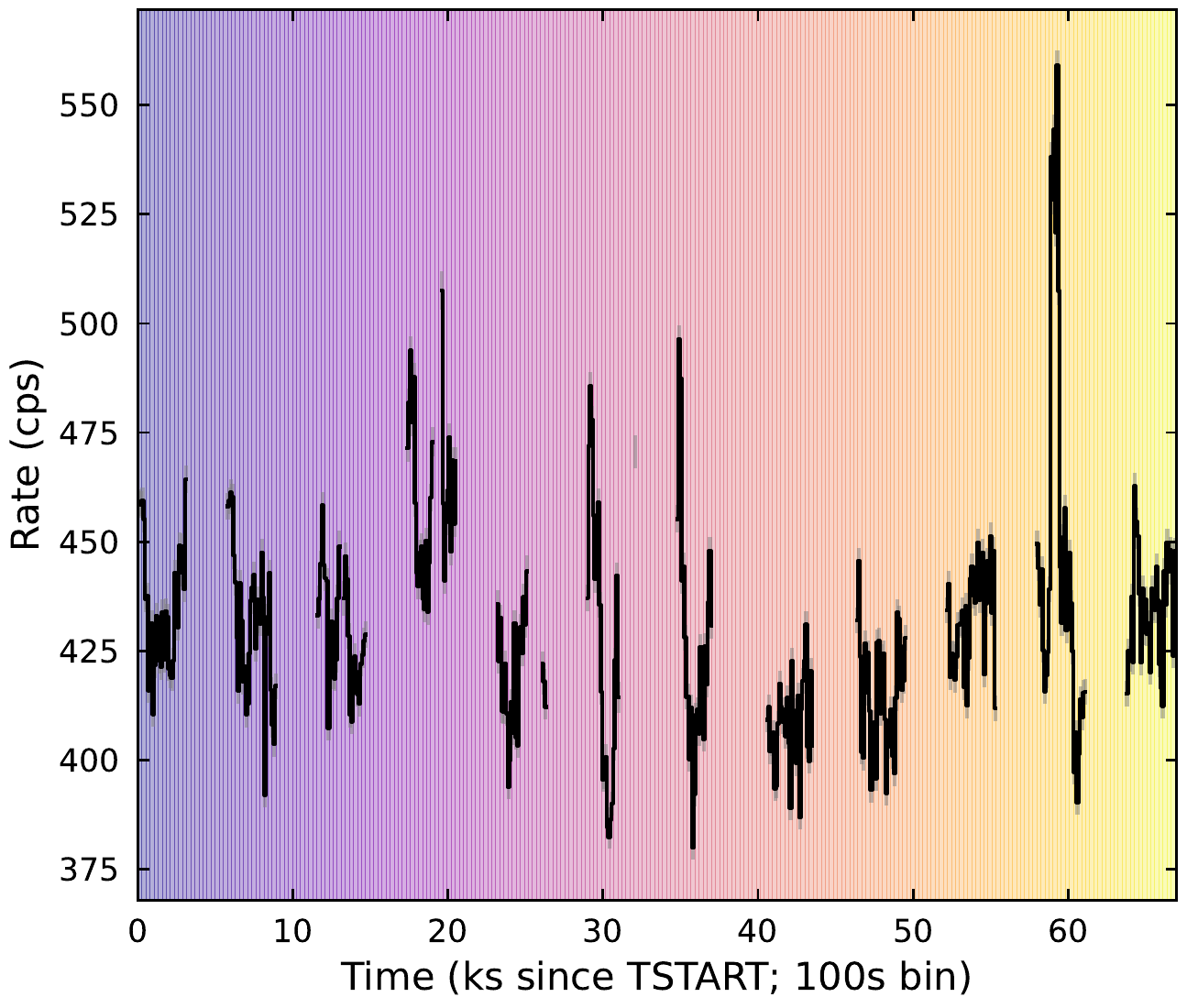}
    \includegraphics[width=\columnwidth]{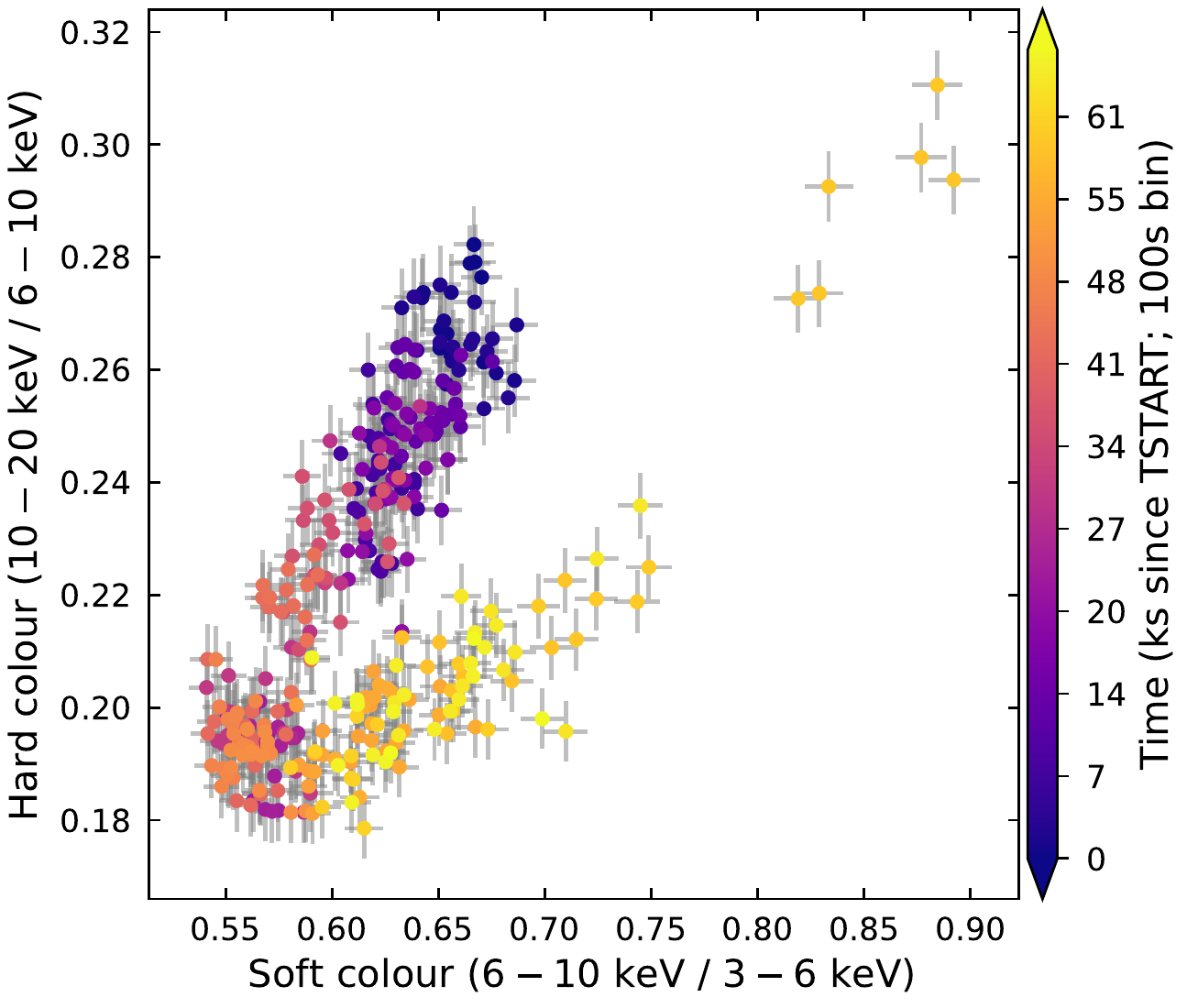}
    \caption{Left panel: {\sl NuSTAR} FPM background-corrected light curves of GX~13+1 with a binning of 100~s, starting at 57851.14686 MJD. The colour gradient of the background of the light curve is associated with the passage of time over the observation. Right panel: colour-colour diagram of GX 13+1 using 100~s bins. The transition from NB to FB can be observed on the lower-left corner of the CCD. We defined $t_{tr}=50$~ks as the transition time between the NB ($t<t_{tr}$) and the FB ($t>t_{tr}$).
    }
    \label{Fig:lctotal}
\end{figure*}

The {\it NuSTAR} telescope \citep[{\sl Nuclear Spectroscopic Telescope Array};][]{2013ApJ...770..103H} is an X-ray satellite equipped with two focal plane modules, FPMA \& FPMB, which are arranged in parallel and contain 2x2 solid-state CdZnTe detectors each, operating in the 3--79~keV energy range. {\it NuSTAR} has a good angular resolution, with a point spread function that varies from 18~arcsec at 10~keV to 58~arcsec at 79~keV. This allows {\it NuSTAR} to accurately locate and study different astronomical objects. The two FPM[A/B] have an energy resolution with a FWHM of 400~eV at 10~keV and 900~eV at 79~keV, enabling {\it NuSTAR} to measure the energies of incoming X-rays with high precision. In terms of sensitivity, {\it NuSTAR} has a 3$\sigma$ sensitivity of 2 $\times$ 10$^{-15}$ erg cm$^{-2}$ s$^{-1}$ in the 6-10 keV range and 10$^{-14}$ erg cm$^{-2}$ s$^{-1}$ in the 10-30 keV range when observing for 1 Ms.

A {\it NuSTAR} observation of GX~13+1 was obtained on April 8, 2017 (ObsID 30301003002), with an effective exposure time of $\sim$24~ks spanning along roughly 69~ks over 12 orbits. Data were reduced using {\tt NuSTARDAS-v.2.0.0} analysis software from the {\tt HEASoft}~v.6.28 software package and {\tt CALDB} V.1.0.2. 
We took the source events that were accumulated within a circular region of 200 arcseconds radius around the focal point. The chosen radius encloses $\sim90$\% of the Point Spread Function (PSF). To collect background events, we selected a circular source-free region with a radius of 150 arcseconds away from the source, on a separate detector. The background average count rate was 4~c~s$^{-1}$ in the energy range of 3--79~keV. We extracted light curves in different energy ranges with different time resolutions.

We used the {\tt nupipeline} task within {\tt NuSTARDAS} to create cleaned event files. Given the high count rate of the source, we set the {\tt statusexpr} keyword as "STATUS==b0000xxx00xxxxxxxx000". To filter out high background activity events from the South Atlantic Anomaly (SAA), we set the corresponding parameters as {\tt saacalc=2}, {\tt saamode=optimized}, {\tt tentacle}=no. This resulted in a loss of approximately $3\%$ of the total exposure time, reducing it from $\sim$24~ks to $\sim$23~ks.
 
 We extracted the light curves and spectra using the {\tt nuproducts} task. We created barycentered lightcurves using the {\tt barycorr} task with the {\tt nuCclock20100101v136} clock correction file. 
 To apply such corrections, we used $(\alpha,\delta)=(273^\circ.6551, -17^\circ.1368)$ and the {\tt JPL-DE2000} ephemeris for GX 13+1. 
 We subtracted the background from each detector and merged the light curves using the {\tt lcmath} task. 

\section{Timing analysis} \label{sec:timing}

In the left panel of \hyperref[Fig:lctotal]{Figure~\ref{Fig:lctotal}} we show a light curve of GX~13+1 with a binning time of 100~s in the 3--79 keV energy range. Different patterns of variability can be observed. In particular, there is a short and bright flare around 61~ks with a peak intensity $\sim$1.4 times the average level of $\sim$415~c~s$^{-1}$. 

We constructed a CCD using background-corrected light curves in the 3--6 keV, 6--10 keV, and 10--20~keV energy bands. 
We defined the hard colour as the ratio between the 10--20 keV to the  6--10 keV light curves, and the soft colour as the ratio between the 3--6 keV and the 6--10~keV rates. We present the resulting diagram in the right panel of \hyperref[Fig:lctotal]{Figure~\ref{Fig:lctotal}}. 

To better represent the temporal evolution across the CCD, we coloured the points in the CCD according to the time from the start of the observation. We found that the source traces a `V' shape path along the CCD, which allowed us to define $t_{tr}=50$~ks as the transition time between the NB ($t<t_{tr}$) and the FB ($t>t_{tr}$).

We searched for the presence of QPOs in the {\it NuSTAR} light curves. We used the Fourier Amplitude Difference (FAD) routines from the {\tt Stingray} package \citep{stingray} to correct for dead time effects. 
We extracted light curves in the intervals 3--6, 6--10, 10--20, and 3--79 keV without subtracting the background, applied the aforementioned routines and obtained the corrected FPM[A/B] power-spectra and FPM[A/B] co-spectra.  We searched for QPOs both in the total observation and in each branch separately in the energy intervals mentioned above. We did not detect any significant QPOs.

\section{Spectral Analysis} \label{sec:spectral}

\begin{figure} 
     \centering
        \includegraphics[width=0.5\textwidth]{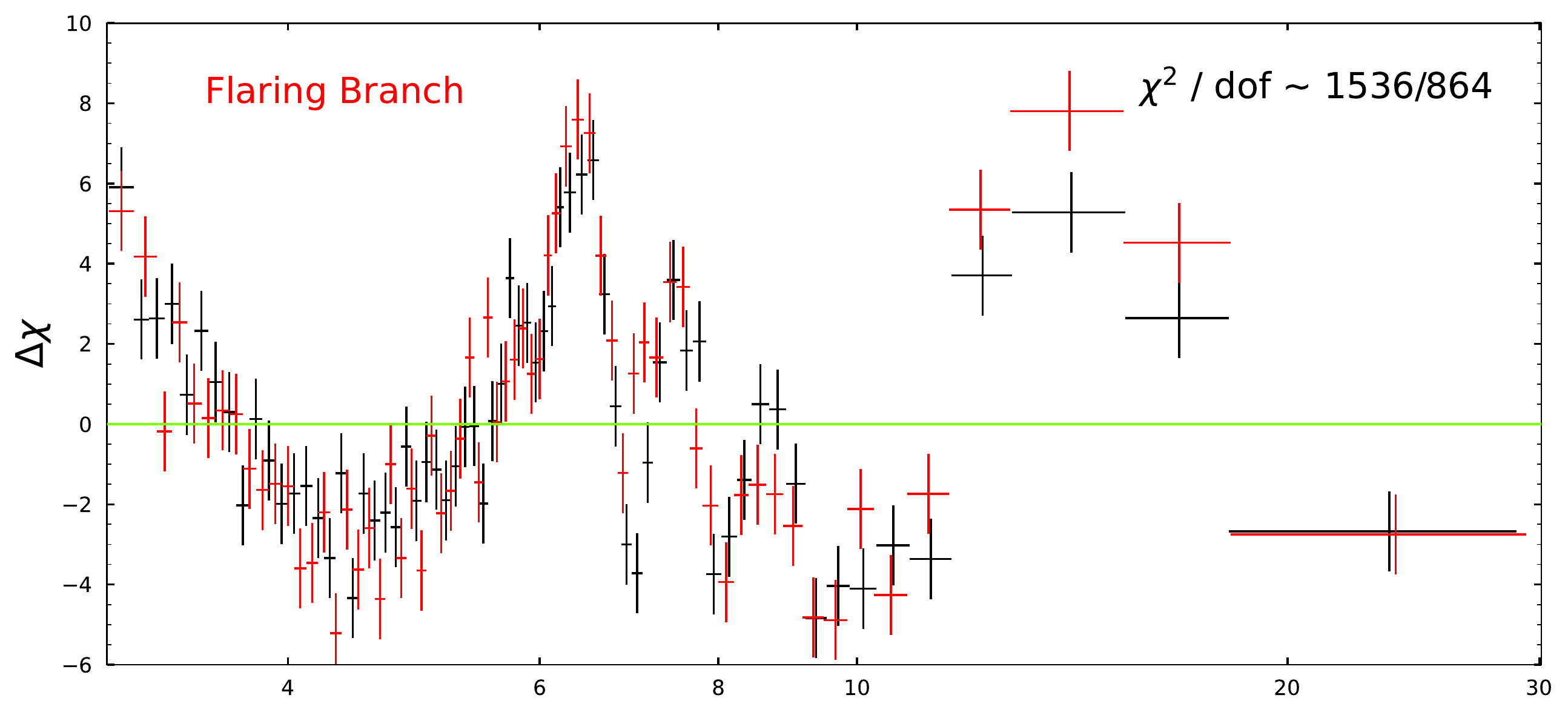}
        \includegraphics[width=0.5\textwidth]{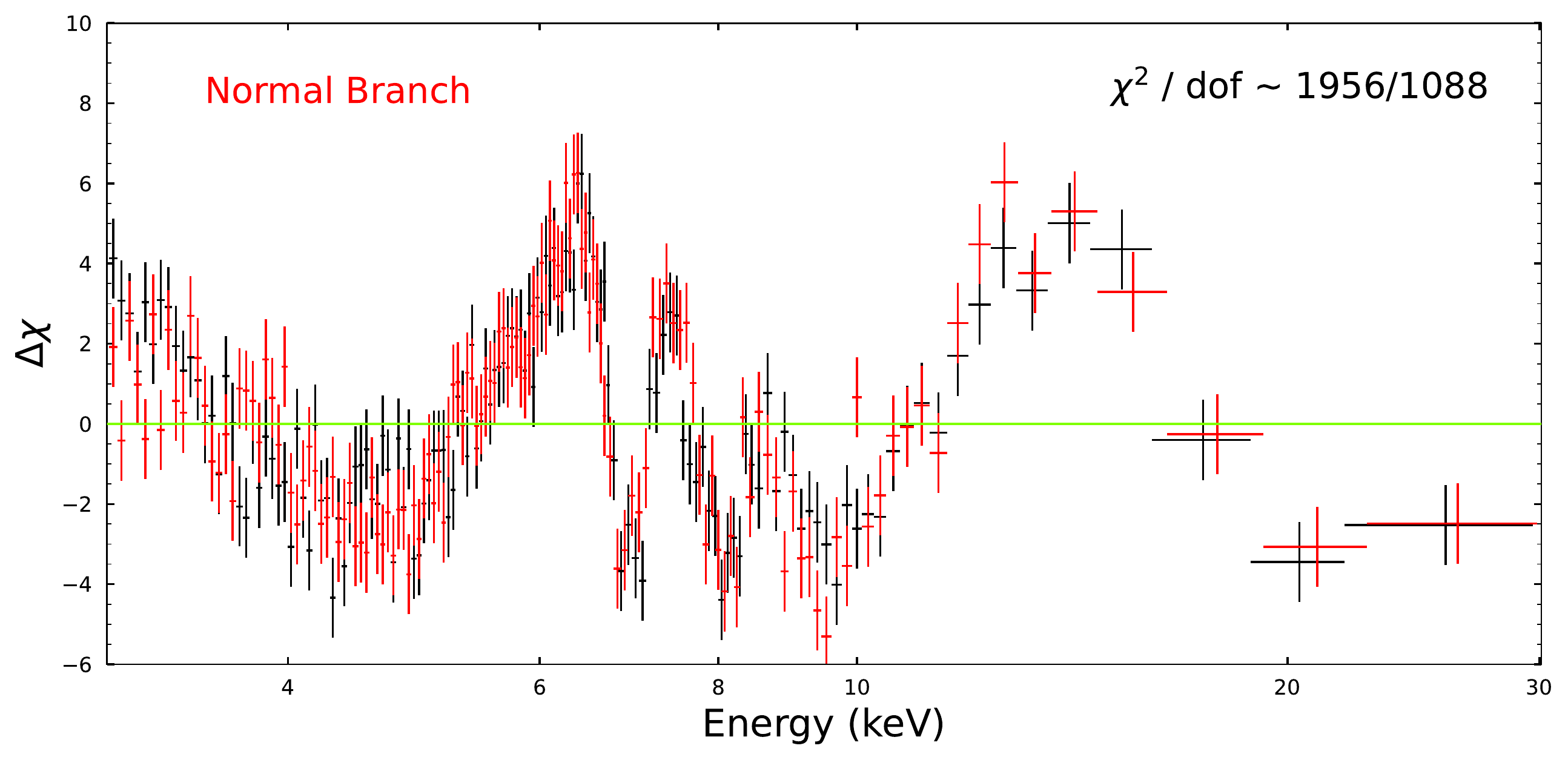}
        \caption{Fit residuals associated with the phenomenological continuum Model 0: 
        {\tt const}*{\tt tbabs}*({\tt bbody} + {\tt diskbb} + {\tt cutoffpl}). The top panel corresponds to the Flaring Branch and the bottom panel corresponds to the Normal Branch. In both panels, red/black residuals correspond to the FPM[A/B] detectors. At the top-right corner of each panel we give the statistics of the represented fit. Although the model used was insufficient to optimise the fit, it was useful to identify the possible absorption features present in the spectra. Relatively-narrow negative residuals are seen around $\sim$7, $\sim$8~keV. The data were re-binned for better visualisation.} 
        \label{fig:residcutoff}
\end{figure}

We used the XSPEC~v12.12.1 package \citep{1996ASPC..101...17A} to model the spectra. We extracted the average spectra for each camera and re-binned the spectra to have at least 25 counts per energy bin in the 3--79 keV energy range in order to apply $\chi^2$ statistics during the fits.

We characterised the spectra of GX 13+1 by performing a broadband spectral analysis, using both FPM[A/B]. As the background became significant at energies above 30 keV, we limited our analysis to the 3--30~keV energy range. In this range, the background flux was 87\% below the source flux. 

We included a multiplicative factor that we fixed to one for the FPMA and left free for the FPMB to account for differences in the effective-area calibration of FPM[A/B]. As previously mentioned in \hyperref[sec:timing]{Section~\ref{sec:timing}}, we divided the total exposure into NB and FB and applied the same spectral model to each branch.

We report all parameter errors at a 90\% confidence level using the Markov Chain Monte Carlo ({\tt chain}) task in {\tt XSPEC}. The chains used the Goodman-Weare algorithm for a total of $5\times10^6$ steps, burning the first $10^6$ and using a total of 200 walkers. To verify the chain convergence we visually inspected that each parameter time series (trace plots) had sufficient state changes, i.e., random-like behaviour. Numerically, we computed the normalised auto-correlation time\footnote{\href{https://emcee.readthedocs.io/en/stable/tutorials/autocorr/}{https://emcee.readthedocs.io/en/stable/tutorials/autocorr/}} and checked that it remained close to unity for each parameter \citep[for more details see][]{Fogantini2022arXiv221009390F}.

\subsection{Continuum emission and reflection}

We used the Tuebingen-Boulder model ({\tt tbabs}) for interstellar absorption, setting solar abundances according to \citet{2000ApJ...542..914W} and using the cross-sections tables given by \citet{1996ApJ...465..487V}. The only free parameter of the {\tt tbabs} component is $N_{\rm H}$, the column density of the absorbing component along the line of sight. 

We firstly fitted the continuum with a combination of blackbody-like components ({\tt bbody} and {\tt diskbb}) and a power law with a cutoff at high energies (Model 0, in XSPEC: {\tt tbabs*(diskbb+bbody+cutoffpl}). For the NB, the best-fitting model ($\chi^{2}_{\nu} = 1.8$) yielded temperatures $kT_{\rm bb} = 1.1 \pm 0.1$ keV and $kT_{\rm in} = 2.8 \pm 0.1$ keV for the blackbody and the accretion disc, respectively. The power law photon index was $\Gamma = -2.5\pm0.1$, and the cutoff energy was 1$\pm$0.5~keV. For the FB, the best-fitting model ($\chi^{2}_{\nu} = 1.7$), with blackbody and accretion disc temperatures of 1.1$\pm$0.1~keV and 1.0$\pm$0.1~keV, respectively. For the power-law component, $\Gamma = -2.5\pm0.1$, and the cutoff energy was 1$\pm$0.5~ke. The residuals of these models are shown in \hyperref[fig:residcutoff]{Figure~\ref{fig:residcutoff}}. We observed several narrow emission and/or absorption features, as well as an indication of an asymmetric Fe~K line profile and a possible Compton hump around 10--25~keV.

\begin{figure}
     \centering
         \includegraphics[width=0.5\textwidth]{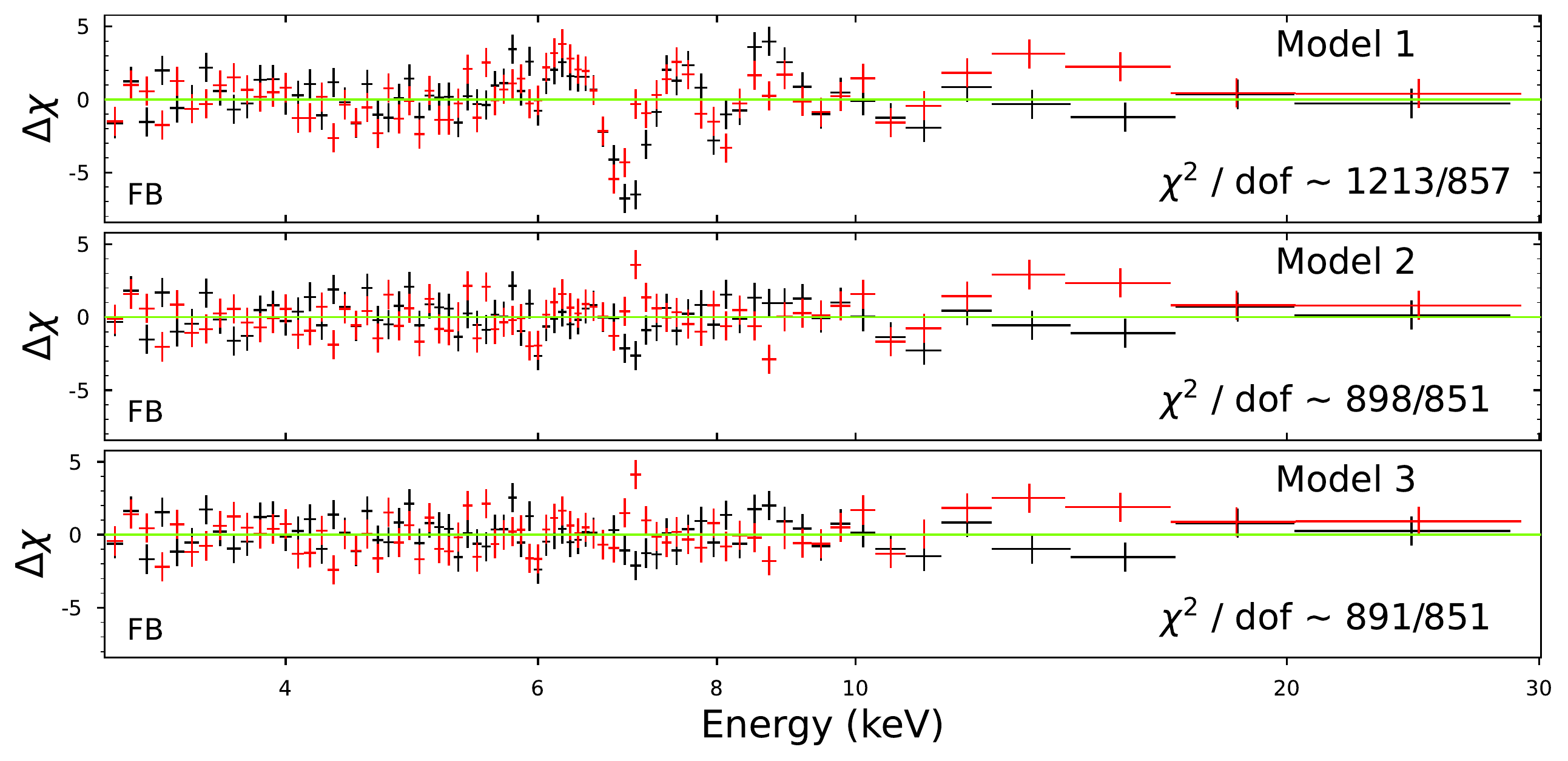}
         \includegraphics[width=0.5\textwidth]{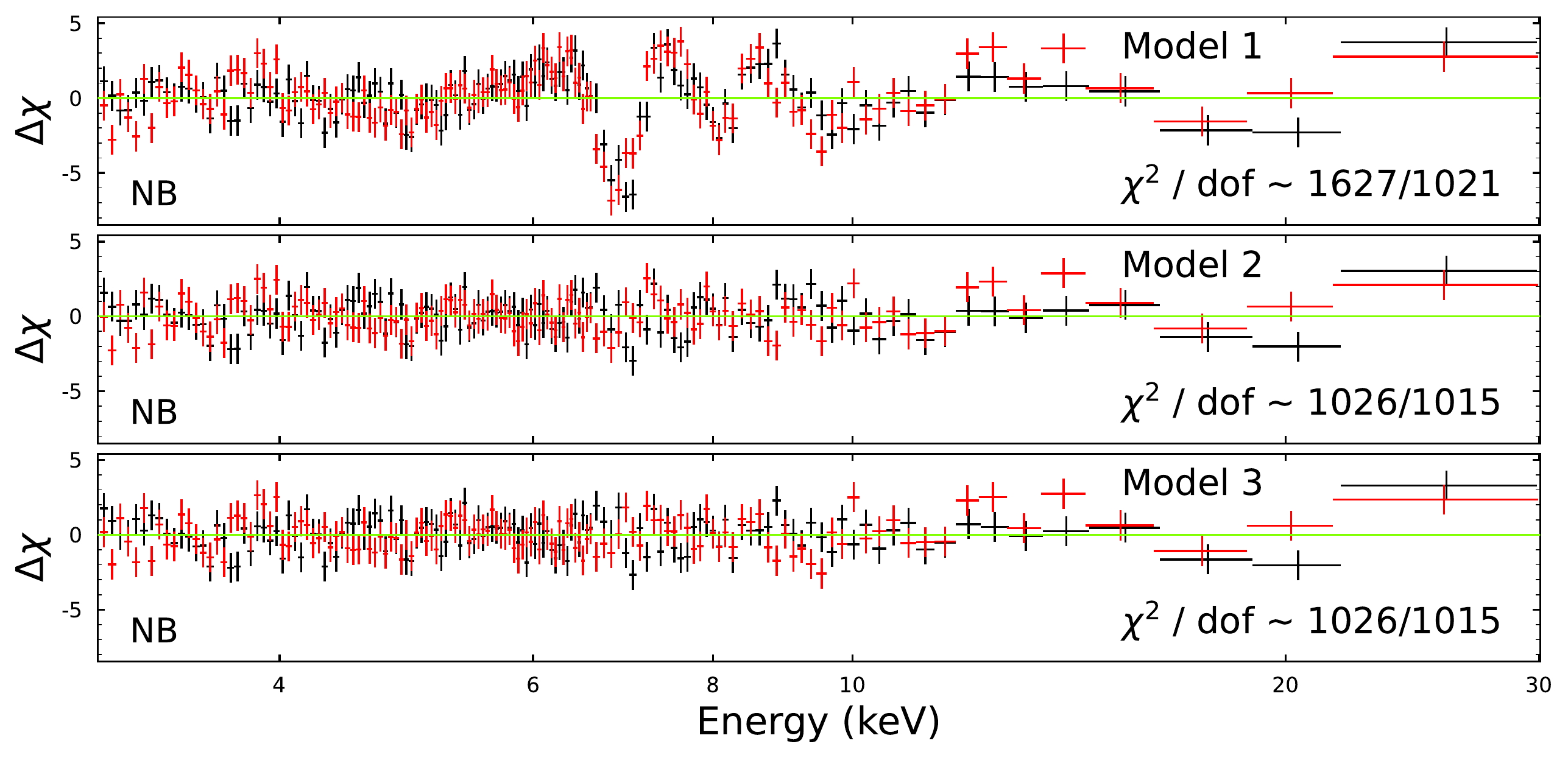}
        \caption{Fit residuals for the fits with Models 1, 2 and 3. The upper panels correspond to the FB and the lower panels to the NB. Red/black residuals correspond to the FPM[A/B] detectors. Model 1 expression is {\tt const}*{\tt tbabs}*({\tt thcomp}*{\tt bbody} + {\tt diskbb} + {\tt relxillNS}, Model 2 expression is {\tt const}*{\tt tbabs}*({\tt thcomp}*({\tt bbody} + {\tt diskbb} + {\tt relxillNS} + {\tt gauss} + {\tt gauss}) and Model 3 expression is {\tt const}*{\tt tbabs}*({\tt thcomp}*{\tt gabs}*{\tt gabs}*({\tt bbody} + {\tt diskbb} + {\tt relxillNS}). The data were re-binned for better visualisation.} 
        \label{fig:resid}
\end{figure}

\begin{table*}
\centering
\resizebox{2\columnwidth}{!}{%
\begin{tabular}{cc|lll|lll}
 \multicolumn{2}{c|}{}           & \multicolumn{3}{c|}{Normal Branch} & \multicolumn{3}{c}{Flaring Branch} \\\hline \hline
Component & Parameters & Model 1   & Model 2   & Model 3  & Model 1    & Model 2   & Model 3  \\ \hline \hline
{\tt constant}  & $C_{\rm FPMB}$& 1.03$\pm$0.001 & 1.03$\pm$0.001 & 1.03$\pm$0.001 & 1.03$\pm$0.001 & 1.03$\pm$0.001 & 1.03$\pm$0.001\\
 {\tt tbabs} & \multicolumn{1}{c|}{$N_{\rm H}~[\rm 10^{22}~cm^{-2}]$} & 2.2$\pm$0.2 & 1.8$\pm$0.3 & 2.0$\pm$0.3 & 2.0$\pm$0.3 & 2.0$\pm$0.4 & 2.0$\pm$0.8 \\
{\tt thcomp} &  \multicolumn{1}{c|}{$\Gamma$}  & 2.3$\pm$0.1 & 2.5$\pm$0.1 & 2.45$\pm$0.1 & 2.1$\pm$0.1 & 2.1$\pm$0.3 & 2.2$\pm$0.1     \\
& \multicolumn{1}{c|}{$kT_{\rm e}~[\rm keV]$} & 2.7$\pm$0.1 & 2.5$\pm$0.1 & 2.9$\pm$0.1 & 2.6$\pm$0.1 & 2.0$\pm$0.1 & 2.6$\pm$0.1 \\
& \multicolumn{1}{c|}{$f_{\rm sc}$} & 0.9$\pm$0.1 & 0.9$\pm$0.1  & 0.9$\pm$0.1 & 0.9$\pm$0.1 & 0.9$\pm$0.1 & 0.9$\pm$0.1 \\
{\tt bbody} & \multicolumn{1}{c|}{$kT_{\rm bb}~[\rm keV]$} & 0.9$\pm$0.1 & 1.15$\pm$ & 1.11$\pm$0.03 & 1.13$\pm$0.1 &  1.1$\pm$0.05 & 1.16$\pm$0.05 \\
&  \multicolumn{1}{c|}{{norm}~$_{\rm bb}$} & 0.06$\pm$0.01 & 0.03$\pm$0.01 & 0.04$\pm$0.01 & 0.04$\pm$0.01 &  0.04$\pm$0.01 
 & 0.04$\pm$0.01  \\
& \multicolumn{1}{c|}{{flux}~$_{\rm bb~compt}$} & 2.8$\pm$0.01 & 2.6$\pm$0.01 & 2.54$\pm$0.01 & 2.93$\pm$0.01 &  2.77$\pm$0.01 & 3.22$\pm$0.01 \\
{\tt diskbb} & \multicolumn{1}{c|}{$kT_{\rm in}~[\rm keV]$} & 0.6$\pm$0.1 &  0.9$\pm$0.1 & 0.7$\pm$0.05 & 0.5$\pm$0.1 &  0.8$\pm$0.05 & 0.6$\pm$0.03 \\
& \multicolumn{1}{c|}{{norm}~$_{\rm dbb}$} & 2965$_{-220}^{+455}$ & 470$\pm$96 & 1508$_{-371}^{+290}$ & 4534$\pm$850  &  560$\pm$163 & 4327$_{-60}^{+200}$ \\
& \multicolumn{1}{c|}{{flux}~$_{\rm dbb}$} & 1.7$\pm$0.01 & 1.3$\pm$0.01 & 1.92$\pm$0.01& 1.32$\pm$0.01 & 1.25$\pm$0.01 &  1.55$\pm$0.01 \\
{\tt relxill$_{\tt NS}$} & \multicolumn{1}{c|}{$q$} & 2.0$\pm$0.1 & 3.4$_{-0.3}^{+0.5}$ & 1.77$\pm$0.25 & 2.2$\pm$0.3 & 2.9$\pm$0.5 & 1.85$\pm$0.26 \\
& \multicolumn{1}{c|}{$\theta~{[\rm deg}]$} & 68$_{-1}^{+6}$ &  27$\pm$2 & 70$_{-2}^{+5}$ & 70$_{-3}^{+5}$ &  29$\pm$2 & 70$\pm$4 \\
&  \multicolumn{1}{c|}{$R_{\rm in}$~[ISCO]} & 1.5$_{-0.2}^{+0.3}$ & 1.4$\pm$0.2 & 1.7$_{-0.69}^{+1.05}$& 2.7$_{-1.4}^{+2.1}$ & 1.4$\pm$0.3 & 1.52$_{-0.43}^{+1.38}$  \\
& \multicolumn{1}{c|}{$\log \xi$~[erg~cm~s$^{-1}$]} & 2.5$\pm$0.1 & 2.5$\pm$0.1 & 2.67$\pm$0.1 & 2.9$\pm$0.2 & 2.9$\pm$0.2 & 2.86$\pm$0.2 \\
& \multicolumn{1}{c|}{$A_{\rm Fe}$} & 0.6$\pm$0.1  & 1.85$^{+0.1}_{-0.4}$ & 1$^{+0.9}_{-0.1}$ & 0.6$\pm$0.1 & 2$\pm$1 & 1$^{+0.2}_{-0.3}$ \\
& \multicolumn{1}{c|}{$\log n_e$~[cm$^{-3}$]} & 15.3$_{-0.2}^{+1.5}$ &  17.2$\pm$0.2 & 16.2$\pm$0.9 & 16$\pm$1 & 17.8$\pm$0.4 & 17.1$\pm$0.9   \\
&  \multicolumn{1}{c|}{{norm}$_{\tt relxill_{NS}}$~[$10^{-2}$]} & 1.7$\pm$0.2 & 1.3$\pm$0.1 & 1.5$\pm$0.2 & 1.5$\pm$0.01 & 1.8$\pm$0.01 & 1.3$\pm$0.2 \\
& \multicolumn{1}{c|}{{flux}~$_{\rm relxill_{NS}}$} & 1.04$\pm$0.01 & 1.8$\pm$0.01 & 0.90$\pm$0.01 & 0.97$\pm$0.01 & 1.85$\pm$0.01 & 1.1$\pm$0.08 \\
{\tt gabs}      &  \multicolumn{1}{c|}{$E$~[keV]}                        & -   & -  & 6.87$\pm$0.02  & -  & - & 6.91$\pm$0.03   \\
                &  \multicolumn{1}{c|}{$\sigma$~[keV]}                   & -   & -  & 0.15$\pm$0.02  & -  & - & 0.14$\pm$0.03   \\
                &  \multicolumn{1}{c|}{{Strength}~[$10^{-2}$]}            & -  & -  & 2.39$\pm$0.8   & -  & - & 5.67$\pm$0.01   \\
{\tt gabs}      &  \multicolumn{1}{c|}{$E$~[keV]}                        & -   & -  & 8.03$\pm$0.05  & -  & - & 8.05$\pm$0.01  \\
                &  \multicolumn{1}{c|}{$\sigma$~[keV]}                    & -  & -  & 0.15$\pm$0.05  & -  & - & 0.13$\pm$0.03  \\
                &  \multicolumn{1}{c|}{{Strength}~[$10^{-2}$]}            & -  & -  & 5.7$\pm$0.4    & -  & - & 3.1$\pm$1.3    \\
{\tt gauss}      &  \multicolumn{1}{c|}{$E_{Ni~K_{\alpha}}$~[keV]}    & -  & 7.46$\pm$0.02 & - & - & 7.5$\pm$0.02 & -    \\
                &  \multicolumn{1}{c|}{$\sigma_{Ni~K_{\alpha}}$~[keV]} & -  &  0.26$\pm$0.02 &  -  & - & 0.26$\pm$0.03  & -   \\
                &  \multicolumn{1}{c|}{{norm}~[$10^{-3}$]}            & -  & 3.77$\pm$0.8   & -  & - & 5.02$\pm$0.01   & - \\
               & \multicolumn{1}{c|}{{flux}~$_{\rm gauss}$}           & -  & 0.04 $\pm$0.01 & - & - & 0.03 $\pm$0.01  & -   \\ 
{\tt gauss}      &  \multicolumn{1}{c|}{$E_{Ni~K_{\beta}}$~[keV]}     & -  & 8.52$\pm$0.05 & -  & - & 8.54$\pm$0.04  & -\\
                &  \multicolumn{1}{c|}{$\sigma_{Ni~K_{\beta}}$~[keV]} & -  & 0.3$\pm$0.05 & -  & - & 0.27$\pm$0.06   & -\\
                &  \multicolumn{1}{c|}{{norm}~[$10^{-3}$]}            & -  & 1.52$\pm$0.4 & -  & - & 2.05$\pm$0.4  & - \\
               & \multicolumn{1}{c|}{{flux}~$_{\rm gauss}$}           & - & 0.02$\pm$0.01 & - & - & 0.02$\pm$0.01 & -  \\  
                \hline
                   \multicolumn{2}{c|}{$\chi^{2}/{\rm dof}~\sim~\chi^{2}_{\nu}$}  & 1627/1021 $\sim$ 1.6 & 1026/1015 $\sim$ 1.01 & 1026/1015 $\sim$ 1.01 & 1213/857 $\sim$ 1.4 & 898/851 $\sim$ 1.05 & 891/851 $\sim$ 1.05 \\ \hline
\end{tabular}%
}
  \caption{Best-fitting parameters derived from the spectral fits for the Normal and Flaring Branches for the following models: \\
  Model 1: {\tt const}*{\tt tbabs}*({\tt thcomp}*{\tt bbody} + {\tt diskbb} + {\tt relxillNS}); \\
  Model 2: {\tt const}*{\tt tbabs}*({\tt thcomp}*{\tt bbody} + {\tt diskbb} + {\tt relxillNS} + {\tt gauss} + {\tt gauss}); \\  
  Model 3: {\tt const}*{\tt tbabs}*{\tt gabs}*{\tt gabs}*({\tt thcomp}*{\tt bbody} + {\tt diskbb} + {\tt relxillNS}). \\
  All fluxes are in the units of 10$^{-9}$~erg~cm$^{-2}$~s$^{-1}$ in the 3--30 keV energy range.}
\label{tab:relxillNS}
\end{table*}

The {\tt bbody} consists of two free parameters, the temperature $kT$ and the normalisation. The {\tt bbody} normalisation is reported in units of ${L_{39}}/{D_{10}^2}$, where $L_{39}$ is the source luminosity in units of $10^{39}$erg~s$^{-1}$ and $D_{10}$ is the distance to the source in units of 10~kpc. The {\tt diskbb} consists of two free parameters, the temperature $kT_{\rm in}$ and the normalisation. The {\tt diskbb} normalisation is reported in units of ${R_{\rm in}}/{D_{10}^2}\cos(\theta)$, where $R_{\rm in}$ is the apparent inner disk radius in km, $D_{10}$ is the distance to the source in units of 10~kpc and $\theta$ is the angle of the disk ($\theta = 0$ is face-on).

We improved the continuum model by replacing the power-law with the {\tt thcomp} thermal Comptonisation component \citep{Zdziarski2020MNRAS.492.5234Z}, which is defined by three parameters: the photon index $\Gamma$, the electron temperature $kT_{\rm e}$, and the scattered/covering fraction $f_{\rm sc}$. $f_{\rm sc}$ ranges from 0 to 1, where a value of 1 indicates that all of the seed photons are Comptonised. We tested two different cases for Comptonisation: one with seed photons originating from the NS ({\tt bbody}) and another with seed photons originating from the accretion disk ({\tt diskbb}). We selected the scenario with the Comptonisation associated with the NS blackbody because the results for the accretion disc Comptonisation gave unusual parameter values, e.g. $\Gamma$ near unity, which did not improve the fitting results. 

Given the hint for a potential relativistic Fe K line complex, and the Compton hump possibly associated to a (relativistic) reflection component, we decided to add a {\tt relxill} component \citep[v2.2,][]{Garcia2013ApJ...768..146G, Dauser2014MNRAS.444L.100D} to our model. For this we used {\tt relxillNS}, a variant that assumes the the illuminating source is the NS \citep[][]{Garcia2022ApJ...926...13G}. We tied the blackbody temperature to the temperature of {\tt relxillNS}. As we consider the illumination spectrum separately, we set refl\_flac = --1. We let the inner disc radius, $R_{\rm in}$, emissivity index $q$, inclination $\theta$, ionisation degree $\xi$, Fe abundance, $A_{\rm Fe}$, and electron density $n_e$ free to vary, and we fix the outer disc radius $R_{\rm out} = 990~r_g$, and the spin parameter $a = 0$ \citep{Galloway2008ApJS..179..360G, Miller2011ApJ...731L...7M, Ludman2018ApJ...858L...5L}. The {\tt relxillNS} normalisation is given in Equation A.1 in \citet{Dauser2016A&A...590A..76D}.

Under such assumptions, we define Model 1, which in XSPEC reads {\tt const}*{\tt tbabs}*({\tt thcomp}*{\tt bbody} + {\tt diskbb} + {\tt relxillNS}). For the NB and FB branches, the best-fitting models yield $\chi^2_\nu = 1.6$, and 1.4, respectively, with $\Gamma \sim 2.2$, $kT_e \sim 2.6$~keV, and $f_{\rm sc} \sim 0.9$, and blackbody and accretion disc temperatures of $\sim$1 and $\sim$0.6~keV, respectively. The reflection model points to an inclination $\theta \approx 60^{\circ}$, ionisation degree $\xi \sim 10^{2.6}$~erg~cm~s$^{-1}$, and $n_e \sim 10^{16}$~cm$^{-3}$. We present the full list of best-fitting parameters for Model 1 in \hyperref[tab:relxillNS]{Table~\ref{tab:relxillNS}}. In \hyperref[fig:resid]{Figure~\ref{fig:resid}} we show the best-fitting residuals, where hints for possible emission/absorption lines are apparent.

We used the {\tt cflux} convolution component to calculate the total unabsorbed flux, which gave a flux of $(5.91 \pm 0.01)\times10^{-9}$~erg~cm$^{-2}$~s$^{-1}$ for the NB and $(5.87 \pm 0.01)\times10^{-9}$~erg~cm$^{-2}$~s$^{-1}$ for the FB, both in the $3-30$ keV energy range.

\subsection{Narrow residuals as emission lines}

We identified possible Fe and Ni emission lines around 6.4 keV, 7.35 keV, and 8.5~keV. We first tried to fit them with Gaussian profiles using the model: {\tt cons * tbabs* ( thcomp * bbody + diskbb + relxillNS + gauss + gauss)}, which we will now refer to as Model 2. We found that the interstellar absorption column values are comparable with those obtained by \citet{Diaztrigo2012A&A...543A..50D}. For both branches, the Comptonisation component yields $\Gamma \sim 2.3$, $kT_e \sim 2.2$~keV, $f_{\sc } \sim 0.9$, and temperatures of the blackbody and the disc of 1.1 and 0.9~keV, respectively. We firstly restricted the inclination to 60--80$^\circ$ \citep{Iaria2009A&A...505.1143I}, however, under such circumstances we were not able to obtain good fits, with $\chi^2_{\nu}>1.3$ (for 1011 d.o.f. for the NB and 850 d.o.f. for the FB), leading to particularly noticeable residuals in the Fe K complex. Therefore, we decided to allow the inclination free to vary, which improved the fit significantly, reaching $\chi^{2}_{\nu} = 1.0$ and 1.07 for the NB and FB, respectively, but yielding relatively low inclinations of $27\pm2^{\circ}$. Furthermore, for both branches, the reflection component yields $\xi \sim 10^{2.5}$~erg~cm~s$^{-1}$, and $n_e \sim 10^{17}$~cm$^{-3}$.

\begin{figure}
     \centering
         \includegraphics[width=0.5\textwidth]{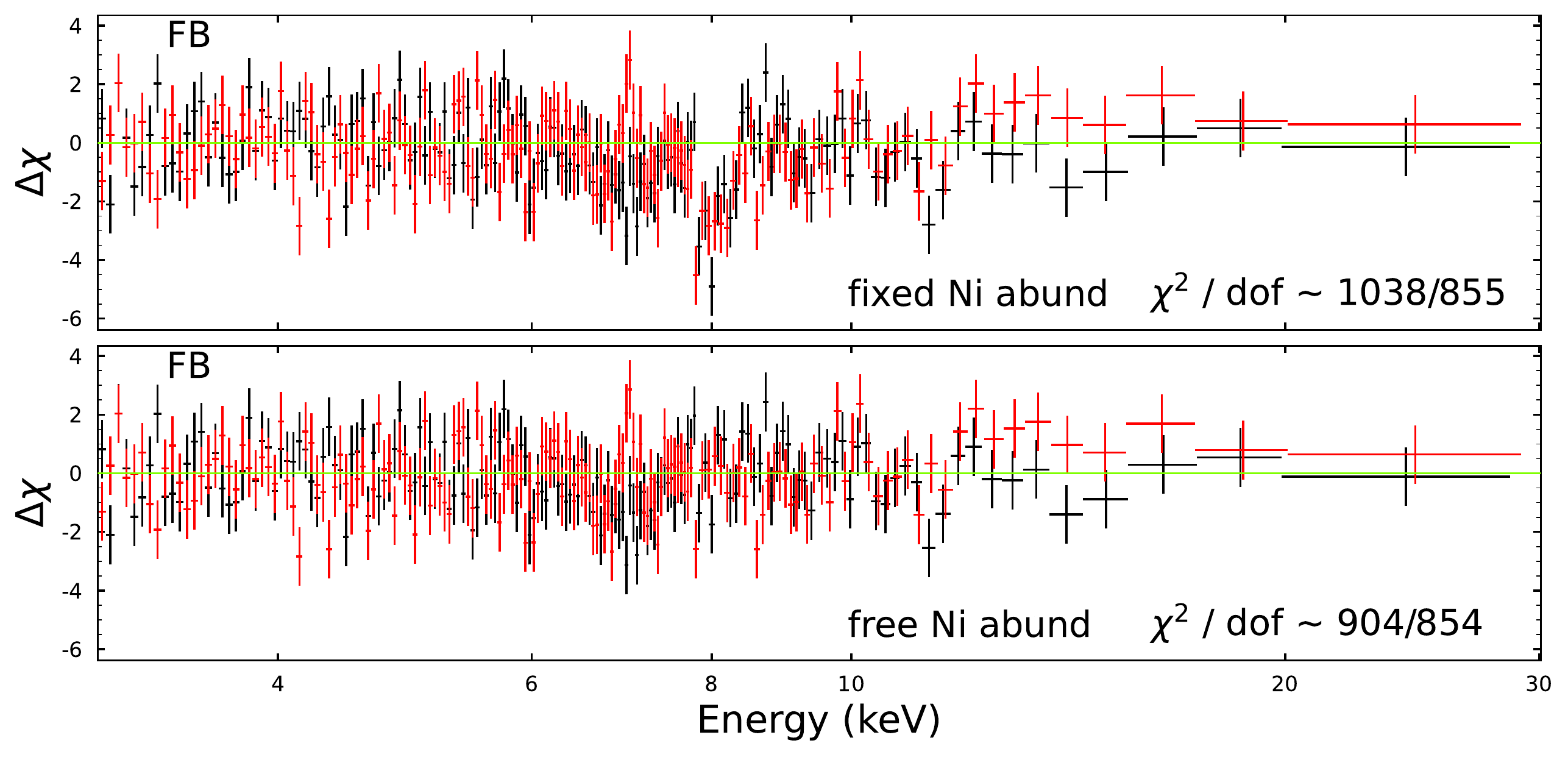}
         \includegraphics[width=0.5\textwidth]{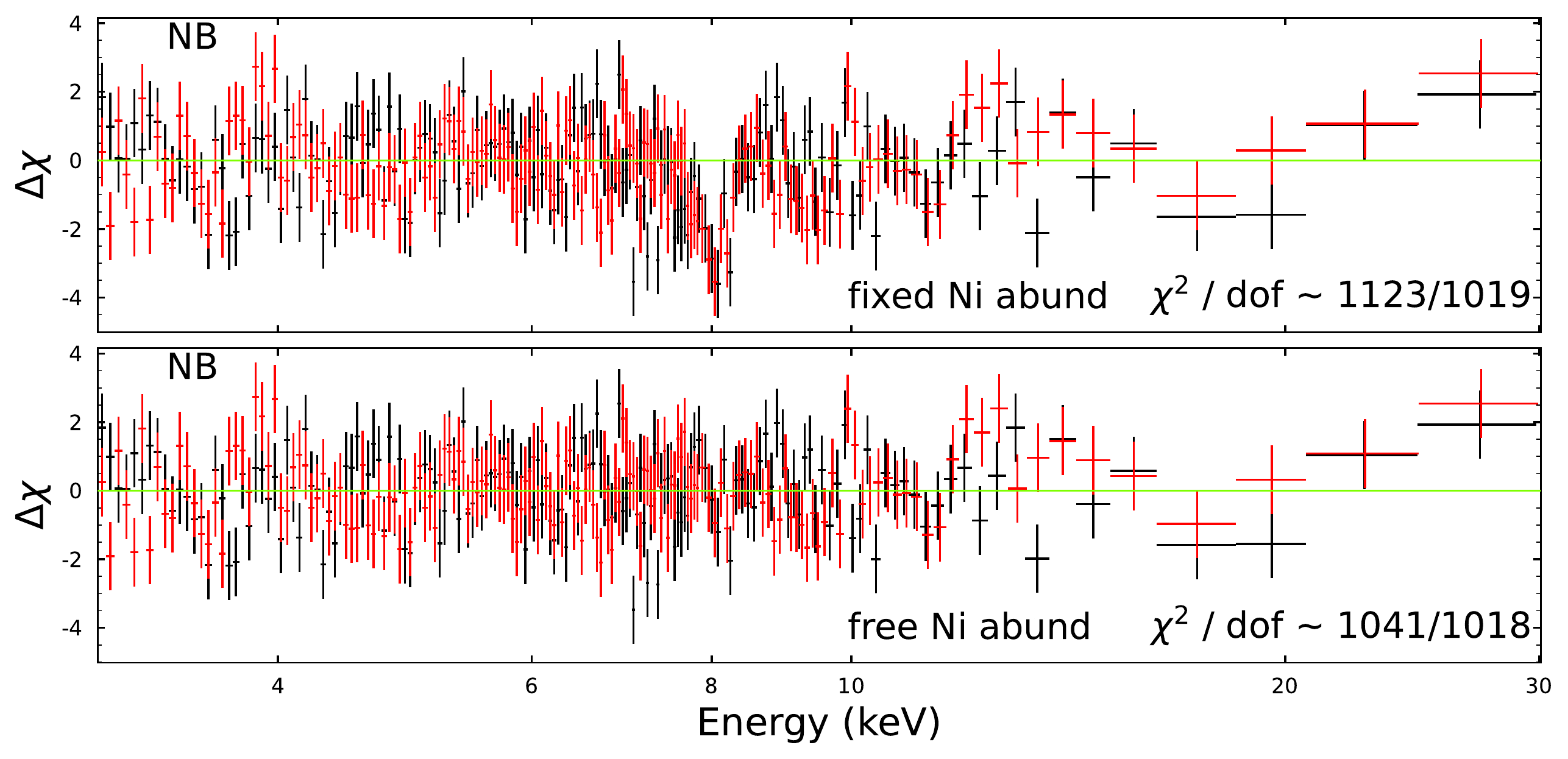}
        \caption{Fit residuals associated with our best-fitting Model: {\tt const}*{\tt tbabs}*{\tt warmabs}*({\tt thcomp}*{\tt bbody} + {\tt diskbb} + {\tt relxillNS}). The top panels correspond to the FB and the bottom panels to the NB. Red/black residuals correspond to FPM[A/B] detectors. The data were re-binned for better visualisation. In both cases, we found this approach to be an excellent fit as well as Model 2. However, the key difference between this approach and Model 1 was the use of a physically motivated model incorporating relativistic reflection and photo-ionised absorption, as opposed to the use of Gaussian components. \\
    {\it Top panels}: Residuals obtained assuming fixed Solar abundances, which fail to fit a narrow absorption feature around 8 keV. \\
    {\it Bottom panels}: Best-fitting residuals obtained leaving the Nickel abundance free to vary, yielding a Nickel overabundance of $\sim$11--12 as reported in Table 2. See text for details}
        \label{fig:warmabas}
\end{figure}

The Gaussian emission lines that we added to the model have central energies of $\approx$ 7.4 and  $\approx$ 8.6~keV, which we identified as putative Ni K${\alpha}$ and K${\beta}$ fluorescence lines, with typical widths $\sigma \approx 0.3$~keV, matching the energy resolution of {\it NuSTAR} at such energies. We further estimated their corresponding equivalent widths (EW) using the {\tt eqwidth} task in {\tt XSPEC}, yielding $148\pm4$ eV and $96\pm5$~eV, for Ni K${\alpha}$ and K${\beta}$, respectively. The final parameters of the model for each branch are listed in \hyperref[tab:relxillNS]{Table~\ref{tab:relxillNS}} and their associated model residuals are shown in \hyperref[fig:resid]{Figure~\ref{fig:resid}}. The {\tt Gaussian} normalisations are reported in units of 10$^{-3}$ photons~cm$^{-2}$~s$^{-1}$ within the line.

 We calculated the total unabsorbed flux using the {\tt cflux} convolution component in XSPEC, which resulted in a flux of $(5.97~\pm0.01)\times10^{-9}$~erg~cm$^{-2}$~s$^{-1}$ for the NB, and $(5.89\pm0.02)\times10^{-9}$~erg~cm$^{-2}$~s$^{-1}$ for the FB, both in the 3--30~keV energy range.

Although, statistically speaking, Model 2 provides excellent spectral fits for both branches, the low inclination derived from this model is not compatible with the presence of as dips in the light curves of the source \citep{2014A&A...561A..99I}. Therefore, we will not consider this scenario further. Moreover, previous studies using other satellites have shown the presence of ionised absorbing material \citep{Diaztrigo2012A&A...543A..50D, Maiolino2019A&A...625A...8M}.

\subsection{Narrow residuals as absorption lines}

 We subsequently modelled the narrow residuals as two Gaussian absorption profiles, {\tt gabs}, which we will refer to as Model 3: {\tt cons * tbabs* gabs * gabs ( thcomp * bbody + diskbb + relxillNS)}. The best-fitting Model 3 for the NB and FB spectra yield $\chi^{2}_{\nu} = 0.99$ and 1.07, respectively. Best-fitting parameters and 90\% error bars are shown in Table~\ref{tab:relxillNS}, and the associated residuals, in Figure~\ref{fig:resid}. 
 
 We found similar absorption column values as with previous models. For both branches, the Comptonisation component yields $\Gamma \sim 2.2$, $kT_e \sim 2-2.5$~keV, $f_{\sc } \sim 0.9$, and temperatures of the blackbody and the disc of 1.1 and 0.6--0.8~keV, respectively.

 The impact of relativistic reflection is more significant in FB than in NB, as shown not only by the flux but also by the residuals shown in \hyperref[fig:residcutoff]{Figure~\ref{fig:residcutoff}}, and in both branches yields a relatively high inclination of $70\pm5^{\circ}$.
 Furthermore, for both branches, the reflection component yields $\xi \sim 10^{2.9}$~erg~cm~s$^{-1}$, and $n_e \sim 10^{16-17}$~cm$^{-3}$.

   \begin{table}
 \centering
 \resizebox{\columnwidth}{!}{%
 \begin{tabular}{@{}clll@{}}
 \toprule
 \multicolumn{4}{c}{Reflection and Photoionised Absorption}                                                                          \\ \hline \hline
 Components        & \multicolumn{1}{l|}{Parameters}                                 & Normal Branch           & Flaring Branch      \\ \hline \hline
 \texttt{constant}  & \multicolumn{1}{l|}{$C_{\rm FPMB}$}                            & 1.03   $\pm$ 0.001 & 1.03  $\pm$ 0.001     \\
 \texttt{tbabs}     & \multicolumn{1}{l|}{$N_{\rm H} [\rm 10^{22}~cm^{-2}]$}         & 2.1    $\pm$ 0.3  & 2.0$_{-1.1}^{+0.3}$  \\
 \texttt{warmabs}   & \multicolumn{1}{l|}{$N_{\rm H, w}~{[\rm 10^{22}~cm^{-2}]}$} &  6$\pm$1.8            & 7.5  $\pm$ 1.7      \\
 \texttt{}          & \multicolumn{1}{l|}{$\log \xi, {\rm w}$~[erg~cm~s$^{-1}$]}     & 3.6   $\pm$ 0.1 & 3.7$\pm$0.1     \\
 \textit{}          & \multicolumn{1}{l|}{$A_{\rm Ni}$}                              & 6   $\pm$ 1  & 5.5 $\pm$ 1.6      \\
 \texttt{}          & \multicolumn{1}{l|}{$\sigma_{v}$~[km~s$^{-1}$]}                & 470$_{-219}^{+451}$  & 730$_{-120}^{+181}$  \\
 \texttt{thcomp}    & \multicolumn{1}{l|}{$\Gamma$}                                 & 2.4    $\pm$ 0.1 & 2.3$\pm$0.1     \\
 \texttt{}          & \multicolumn{1}{l|}{$kT_{\rm e}~[\rm keV]$}                   & 2.9    $\pm$ 0.1 & 2.6$\pm$0.1     \\
 \texttt{}          & \multicolumn{1}{l|}{$f_{\rm sc}$}                             & 0.9    $\pm$ 0.1 & 0.9   $\pm$ 0.1     \\
 \texttt{bbodyrad}     & \multicolumn{1}{l|}{$kT_{\rm bb}~[\rm keV]$}                  & 1.1    $\pm$ 0.1 & 1.1  $\pm$ 0.1     \\
 \texttt{}          & \multicolumn{1}{l|}{{norm}~$_{\rm bb}$~}                     & 274    $\pm$ 4 & 211   $\pm$ 5     \\
 \texttt{}          & \multicolumn{1}{l|}{{flux}~$_{\rm bb~compt}$}                 & 3.05   $\pm$ 0.01 & 3.21  $\pm$ 0.01      \\
 \texttt{diskbb}    & \multicolumn{1}{l|}{$kT_{in}~[\rm keV]$}                      & 0.7    $\pm$ 0.1  & 0.6   $\pm$ 0.1     \\
 \texttt{}          & \multicolumn{1}{l|}{{norm}~$_{\rm dbb}$}                      & 1800    $\pm$ 250  & 5700 $\pm$ 1150      \\
 \texttt{}          & \multicolumn{1}{l|}{{flux}~$_{\rm dbb}$}                      & 1.42   $\pm$ 0.01 & 1.15  $\pm$ 0.01     \\
 \texttt{relxill$_{\tt NS}$} & \multicolumn{1}{l|}{$q$}                             & 1.6   $\pm$ 0.2   & 1.6    $\pm$ 0.3      \\
  \texttt{} & \multicolumn{1}{l|}{$\theta~{[\rm deg}]$}                             & 68   $\pm$ 3  & 67$_{-1}^{+5}$      \\
 \textit{}          & \multicolumn{1}{l|}{$R_{\rm in}$ [ISCO]}                      & <1.6  & <1.8 \\
 \texttt{}          & \multicolumn{1}{l|}{$\log \xi$~[erg~cm~s$^{-1}$]}             & 2.3   $\pm$ 0.1 & 2.4$\pm$0.1      \\
 \textit{}          & \multicolumn{1}{l|}{$A_{\rm Fe}$}                             & 1.4    $\pm$ 0.3  & 1.6$_{-0.5}^{+0.7}$      \\
 \textit{}          & \multicolumn{1}{l|}{$\log n_e$ [cm$^{-3}$]}                   & 19$^\dagger$  & 19$^\dagger$  \\
 \textit{}          & \multicolumn{1}{l|}{{\it norm}$_{{\tt relxill}_{NS}}$~[$10^{-2}$]}   & 1.6   $\pm$ 0.3 & 6.4  $\pm$ 0.4      \\
 \texttt{}          & \multicolumn{1}{l|}{{flux}~$_{\rm relxill_{NS}}$}                  & 1.49   $\pm$ 0.01 & 1.61   $\pm$ 0.01     \\ \hline 
 \multicolumn{2}{c|}{$\chi^{2}/{\rm dof}~\sim~\chi^{2}_{\nu}$}                       & 1041/1018 $\sim$ 1.02  & 904/854 $\sim$ 1.05  \\ \hline
 \end{tabular}
 } 
 \caption{Best-fitting parameters obtained from our model including relativistic reflection {\tt rexillNS} and photo-ionised absorption {\tt warmabs}, for the NB and FB branches.  The $^\dagger$~symbol indicates that the corresponding parameter was frozen. All fluxes are in the units of 10$^{-9}$~erg~cm$^{-2}$~s$^{-1}$ in the 3--30 keV energy range. For more details refer to the text.}
 \label{tab:warmabs}
 \end{table}

 The presence of possible absorption lines is motivated by the higher energy resolution XMM-{\it Newton} observations analysed by \citet{Diaztrigo2012A&A...543A..50D}. We identified two narrow absorption lines at around 6.8 keV as Fe XXVI, and at 8 keV as possibly Ni XXVII, Fe XXV K$\alpha$ or Ni XXVIII, or a combination of these transitions. 

The EW of the 6.8 keV Fe line were 70$\pm$3~eV and 95$\pm$6~eV in the FB and NB, respectively, and for the 8~keV Ni line 54$\pm$4~eV and 45$\pm$3~eV in the FB and NB, respectively.

We compared these results to those obtained by \citet{Diaztrigo2012A&A...543A..50D} and found that the line around 6.8 keV had an approximately 4.5 times higher EW in the NB and approximately 3.4 times higher EW in the FB. Comparing the line around 8 keV was more challenging because it was unclear which transition was involved. For example, when we assumed the transition to be Fe XXVI, we found that our EW was approximately 1.2 times higher in the NB and 2.5 times higher in the FB.
 
 We also calculated the total unabsorbed flux using the {\tt cflux} convolution component in XSPEC, which resulted in a flux of $(5.95~\pm0.01)\times10^{-9}$~erg~cm$^{-2}$~s$^{-1}$for the NB, and $(5.87\pm0.01)\times10^{-9}$~erg~cm$^{-2}$~s$^{-1}$ for the FB, both in the 3--30 keV energy range.

 \subsection{Photo-ionised absorption: {\tt warmabs}}

Considering that the absorption lines that we fitted with {\tt gabs} functions in the previous Section are due to a photo-ionised plasma, we replaced the narrow Gaussian absorption profiles with a physically-motivated component, {\tt warmabs} \citep[v2.47][]{Kallman1996ApJ...465..994K, Kallman2001ApJS..133..221K, Kallamn2004ApJS..155..675K, Kallman2009ApJ...701..865K}. This model allows us to consider the absorption due to a photo-ionised plasma along the line of sight, possibly originating from an ionised wind on top of the accretion disc. The {\tt warmabs} component self-consistently considers the physics of absorption modifying the continuum by introducing absorption features near 7~keV, like those described in the previous subsection. The new model expression in XSPEC reads: {\tt const}*{\tt warmabs}*{\tt tbabs}*({\tt thcomp}*{\tt bbodyrad} + {\tt diskbb} + {\tt relxillNS}). In both branches (FB and NB) of our analysis, we found that it was not possible to fit the feature at $\sim$8~keV if we kept the Ni/Fe abundance in {\tt warmabs}, fixed to the Solar value. Therefore, we decided to leave the Ni abundance, $A_{\rm Ni}$, free to vary during the fit, which allowed us to fit all the absorption features simultaneously with two parameters less than in Model~3. 

The NB yields a turbulent broadening of 470$_{-219}^{+451}$~km~s$^{-1}$, while the FB yields a turbulent broadening of 730$_{-120}^{+181}$~km~s$^{-1}$. The large errors associated with these values are due to the resolution limitations of {\sl NuSTAR}. During the transition from the NB to FB, there were no significant changes in the remaining parameters of the total model. The ionisation parameter remained consistent with $\sim 10^{3.7}$~erg~cm~s$^{-1}$, with a wind absorption column of $\sim 8 \times 10^{22}$~cm$^{-2}$, and a Ni/Fe overabundance of $\sim$6 times solar. Both black bodies and the Comptonisation component yielded values fully compatible with those from the previous phenomenological Models, and the relativistic reflection {\tt relxillNS} model yielded an inclination of $68\pm3^{\circ}$, with an emissivity index of $q \sim 1.7$, $R_{\rm in} < 8.7~r_g$, an ionisation degree $\xi \sim 10^{2.3}$~erg~cm~s$^{-1}$, and electron densities of $n_e \sim 10^{17}$~cm$^{-3}$.

As we will explain in more detail in the Discussion, the best-fitting Model described above, with $n_e \approx 10^{17}$~cm$^{-3}$, leads to problematically large distances for the illuminating source for reflection, which is physically inconsistent given the relatively low ionisation degrees in the source. Since {\tt RelxillNS}, the most complete model available for relativistic reflection, only admits electron densities up to 10$^{19}$~cm$^{-3}$, we thus fixed $n_e$ at this value and tried to fit again the model. As expected, we found that this fit is slightly worse than the one previously described, i.e. leaving $n_e$ free. Quantitatively, for the NB we obtained a $\chi^{2}/{\rm dof}$: 1025/1017 and 1041/1018 for free $n_e$ and fixed $n_e$ at 10$^{19}$~cm$^{-3}$, respectively. Similarly, for the FB we obtained a 899/853 and 904/854 for free $n_e$ and fixed $n_e$ at 10$^{19}$~cm$^{-3}$, respectively. Besides the relatively small change of the $\chi^2$, the residuals associated with these fits look unbiased, and the best-fitting parameters found are consistent with those obtained when $n_e$ is free; we therefore adopt this as our {\em Preferred} Model. Further details for this decision will be provided in the next Section.

In \hyperref[tab:relxillNS]{Table~\ref{tab:relxillNS}} we present our best-fitting parameters and associated 90\% uncertainties for both the NB and FB, and the model residuals are presented in \hyperref[fig:warmabas]{Figure~\ref{fig:warmabas}}. All parameter errors were obtained using the MCMC Goodman-Weare algorithm, with a total of 10$^{5}$ steps and 200 walkers, after burning the first 30000 steps. We obtained a reduced $\chi^{2}$ for the NB and for the FB that is identical to Model~3, but with two fewer free parameters. The most important difference in both cases is that we used a physical model with relativistic reflection and photo-ionised absorption instead of Gaussian functions. 

We calculated the total unabsorbed flux for this preferred Model, obtaining fluxes of $(5.95~\pm0.01)\times10^{-9}$~erg~cm$^{-2}$~s$^{-1}$ for the NB, and $(5.87\pm0.01)\times10^{-9}$~erg~cm$^{-2}$~s$^{-1}$ for the FB, both in the 3--30 keV energy range.

\section{Discussion} \label{sec:Discussion}

We analysed a {\it NuSTAR} observation of the NS-LMXB GX 13+1 at orbital phases that span from 0.55 to 0.6, when the source is not expected to show dips in its light curve. The source transitioned from the Normal to the Flaring Branch during the observation. We obtained energy spectra from both branches and identified a reflection component, evidenced by a Fe-line complex displaying hallmarks of both relativistic and Doppler broadening effects, as well as a prominent high-energy Compton hump. Additionally, we uncovered evidence of narrow residuals, which we attributed to a photo-ionised plasma emanating from the accretion disc. This ionised plasma was found to possess a Ni abundance that is 6 times solar.

We fitted the broad-band spectra of the source in the source in the two branches with a model consisting of a multi-temperature blackbody representing the accretion disc, a thermally Comptonized blackbody for the NS and boundary layer and a relativistic-reflection ({\tt relxillNS}) absorbed by a photo-ionised plasma ({\tt warmabs}). The best-fitting model gives a relatively high inclination $\sim$70$^\circ$, compatible with the dipping features previously observed in the light curves of GX~13+1. We found that the spectral changes from the NB to the FB are very slight, which is a common behaviour in other $Z$-sources \citep[see for e.g.,][]{Daia2009ApJ...693L...1D, Lavagetto2008A&A...478..181L, Agrawal2009MNRAS.398.1352A, Lin2009ApJ...699...60L}. 

The unabsorbed luminosity in the 3--30~keV energy band is approximately $(3.5~\pm~0.1) \times 10^{37}$~erg~s$^{-1}$ for the NB and approximately $(3.44~\pm~0.1) \times 10^{37}$~erg~s$^{-1}$ for the FB, assuming a distance of 7~kpc. Essentially the luminosity does not change significantly within errors at the branch transition. Those values correspond to $\sim$ 9\% of the Eddington luminosity, which is $\sim$ $3.8 \times 10^{38}$~erg~s$^{-1}$ for a canonical 1.4~M$_\odot$ NS \citep{Kuulkers2003A&A...399..663K}. {\it NuSTAR} has observed some of the highest luminosities of LMXBs, including GX~13+1 (this work), Serpens~X-1  \citep[{\it Atoll} source, $\sim$23\%~$L_{\rm Edd}$,][]{Mondal2020MNRAS.494.3177M} and Aquila X-1 \citep[{\it Atoll} source, $\sim$32\%~$L_{\rm Edd}$,][]{Ludlam2017ApJ...847..135L}. \citet{Homan2018ApJ...853..157H} compared the luminosity of GX~5--1 with other LMXBs observed by {\it NuSTAR}, and identified a critical value for the presence of reflection features at approximately 2\% of the Eddington luminosity.

\subsection{Emission or absorption lines and photo-ionisation}

When we fitted the narrow residuals with Gaussian emission lines (Model~2), we found that the relativistic Fe-line complex could only be explained if the source inclination was low, $i~\lesssim$30$^\circ$. A low inclination is inconsistent with previous spectral results obtained with other instruments with better energy resolution than {\it NuSTAR} \citep{Diaztrigo2012A&A...543A..50D, Maiolino2019A&A...625A...8M}. Additionally, a high inclination is necessary to explain the presence of dips in the light curves of GX~13+1 \citep{2014A&A...561A..99I}. Therefore, we ruled out this scenario and decided to try a new model considering the narrow residuals as absorption line features in the so-called Model~3. It should be noted that, while this model gives statistically a good fit, this only highlights the need for extreme caution when working with medium-high resolution data, such as those from {\it NuSTAR}. Model and parameter degeneracy may result in severely biased results, leading to very different conclusions regarding critical parameters, such as system inclination, when analysing a new source.

In Model~3, we considered a similar continuum model as Model~2, but we used multiplicative Gaussian absorption profiles, {\tt gabs}, to fit the narrow residuals that we identified as Fe XXVI K$\alpha$ at $\sim$6.7~keV, and either Ni XXVII and/or Ni XXVIII at $\sim$8~keV. 
Interestingly, our best-fitting Model 3 yielded an inclination of approximately $\sim$70$^\circ$, consistent with previous observations \citep{Diaztrigo2012A&A...543A..50D, Maiolino2019A&A...625A...8M}.

Using the {\tt warmabs} component we found, both in NB and FB spectra, an excess of Ni/Fe\,$\sim$\,6. This striking overabundance bears resemblance to the Ni/Fe ratio observed in the microquasar SS~433. The evidence suggests that this excess may originate from the powerful jets or winds found in high-luminosity sources, as previously proposed by \citet{Medvedev2018AstL...44..390M} and \citet{Fogantini2022arXiv221009390F}. The velocity we measured is in agreement with the findings of \citet{Ueda2004ApJ...609..325U, Diaztrigo2012A&A...543A..50D}. However, it is important to note that {\it NuSTAR} lacks the sufficient spectral resolution to detect smaller dispersion velocities, as well as redshift/blueshift of such lines at non-relativistic speeds.

Following \citet{Diaztrigo2012A&A...543A..50D}, we can use the spectral-fitting results to estimate the distance $r$, between the ionising source and the slab of ionised absorbing material, using the equation:

\begin{equation}
r = \frac{L}{\xi~N_{\rm H, warmabs}}\frac{d}{r},
\end{equation}

where $\xi$ is the ionisation parameter, $n_e \approx N_{\rm H, warmabs}/{d}$ is the electron density of the ionised plasma, and $d$ is the thickness of the slab of ionised absorbing material. Considering typical values of $d/r$ ranging from 0.1 to 1, we estimate $r$ for the NB and FB was between $\sim$~$4 \times (10^{5}$ -- $10^{6}$)~km. These values are in agreement with those previously found by \citet{Diaztrigo2012A&A...543A..50D} and \citet{Ueda2004ApJ...609..325U} for GX~13+1 using XMM-{\it Newton} and {\it Chandra} data, respectively. 

\subsection{Inner disk radii and magnetic field strength} 

 The {\tt relxillNS} model in combination with the {\tt warmabs} model can be used to determine the inner radius of the accretion disc. In the case of the NB, the inner radius was found to be < 9.6 $r_{g}$ $\approx$ 19.2~km and for the FB, the inner radius was found to be < 10.8 $r_{g}$ $\approx$ 21.8~km, assuming in both cases a mass of 1.4M$_\odot$. We thus find no significant differences in the inner disk radii of the two states within the associated errors. This is consistent with the results of previous studies that have examined how the line profile changes with the source state during an observation of a $Z$-source \citep{Daia2009ApJ...693L...1D, Iaria2009A&A...505.1143I}.

We can set an upper limit on the equatorial magnetic field strength of the NS by adopting equation (15) from \citet{Ibragimov2009MNRAS.400..492I} for the magnetic dipole moment and using the upper limit of $R_{\rm in}$ measured from the reflection fit:

\begin{equation}
\begin{split}
    B = 2.4~\times~10^7~k_{\mathrm{A}}^{-7/4}~\left( \frac{M}{1.4M_{\odot}}\right)^2 \left( \frac{10~ 
    \rm km}{{R_{\rm NS}}}\right)^3~\left( \frac{{R_{\rm in}}}{10~\rm km}\right)^{7/4} \\
    \times~\left( \frac{{\it f_{\rm ang}}}{\eta} \frac{{\it F_{\rm bol}}}{10^{-9}~\mathrm{erg~cm^{-2}~s^{-1}}}\right)^{1/2} \frac{D}{15~{\rm kpc}}~\mathrm{G}.
\end{split}
\end{equation}

We assumed a distance of 7~kpc, an accretion efficiency $\eta = 0.2$, as reported in \citet{Sibgatullin2000AstL...26..699S}, a conversion factor $k_{A}$ $\approx$ 1 \citep{Ibragimov2009MNRAS.400..492I}, an angular anisotropy ($f_{\rm ang}$) of approximately unity \citep{Ludlam2019ApJ...873...99L}, and a canonical mass of 1.4~M$_\odot$. Using the $R_{\rm in}$ value found in FB, we obtained a magnetic field strength of $B \lesssim 1.8 \times 10^8$~G, where the upper bound considers both the FB and the NB. This result is consistent with the magnetic field strengths of other accreting NS-LMXBs \citep{Cackett2009ApJ...694L..21C, Mukherjee10.1093/mnras/stv1542, Ludlam2017ApJ...847..135L, Ludlam2019ApJ...873...99L, Ludlam2020ApJ...895...45L, Ludlam2021ApJ...911..123L, Pan2018MNRAS.480..692P, Sharma2020MNRAS.496..197S, Mondal2022MNRAS.516.1256M}.

Determining the upper limit of the magnetic field in a $Z$-source is important because it helps us to better understand the physical processes underlying the phenomena observed in these sources. In doing so, it can help to better understand the mechanisms governing the behaviour of $Z$-sources and to place limits on models of magnetic field generation and evolution in NSs \citep{Zhang_10.1007/978-3-540-74713-0_114, DiSalvo2022ASSL..465...87D}.

Previous studies have suggested that the magnetic field strength of $Z$-sources is larger than that of {\it Atoll} sources \citep{focke1996ApJ...470L.127F, Zhang2006MNRAS.366..137Z, Chen2006ApJ...650..299C, Ding2011AJ....142...34D} and millisecond X-ray pulsars \citep{Cackett2009ApJ...694L..21C, Disalvo2003A&A...397..723D}. 
We have never observed pulsations in $Z$-sources, despite having a magnetic field strength that is similar to or higher than other sources mentioned above. \citet{Pringle1972A&A....21....1P} suggested that the emission of pulsations from a NS may depend on the shape of the emission cone, the orientations of the magnetic and rotation axes, and the line of sight. \citet{Lamb1973ApJ...184..271L} proposed that the alignment of the NS magnetic axis with the axis of rotation may influence the detection of pulsations. Therefore, it is possible that the NS magnetic axis in $Z$-sources is parallel to the axis of rotation, or that the orientation of the rotation axis is favourable for observation, leading to the absence of pulsations in these sources. This is consistent with what was observed in GX 13+1. 

\subsection{The boundary layer illuminating the disk}

The boundary layer is the region between the NS surface and the position where the angular velocity of the accretion disk is maximum. When the accreting gas in this region decelerates, a significant amount of kinetic energy must be converted into heat, thus emitting radiation. The amount of energy lost depends on the rotational velocity of the NS. Our fit shows that the Comptonisation component associated with this region dominates the spectra from 7 to 30~keV, similar to other LMXBs, except for the accreting millisecond pulsars \citep{Barret2000ApJ...533..329B}. This Comptonised blackbody component may originate in such a BL region \citep[see e.g.,][]{Inogamov1999AstL...25..269I, Popham2001ApJ...547..355P, Grebenev2002AstL...28..150G, Gilfanov2003A&A...410..217G, Revnivtsev2006A&A...453..253R}. The dominance of the Comptonised blackbody component in the spectra suggests that the BL may be the main source of ionising flux illuminating the inner accretion disc, which eventually leads to the observed Fe~K complex emission.

To determine the maximum radial extent of a boundary layer from the surface of the NS based on the mass accretion rate, we used Equation 25 from \citet{Popham2001ApJ...547..355P}. We note that this equation does not take into account the size of the boundary layer in the vertical direction or the impact of the NS rotation. We calculated the associated accretion rate using the luminosity, which yields $3 \times 10^{-9}$~M$_{\odot}$~yr$^{-1}$, and obtained a BL with a radial extent of 3~km from the NS surface.

The {\tt bbodyrad} normalisation is $R_{\rm km}^2/D_{10}^2$, where $R_{\rm km}$ is the radius of the source in km and $D_{10}$ is the distance to the source in units of 10 kpc. The associated spherical emission has a radius between 10--12~km assuming a distance of 7~kpc. Using the radial extent of the BL of 3~km, and the inner-disc radius of $R_{{\rm in}} \approx 19$~km, we can constrain the NS radius, $R_{\rm NS}$, to be less than $\sim$16~km.

Using the reflection parameters, we can obtain the density of the accretion disc. Following \citet{Cackett2010ApJ...720..205C}, the ionisation degree in the inner accretion disk, $\xi$, is calculated as $ \xi = \frac{L_{BL}}{n_e~R^2}$ where $L_{\rm BL}$ is the BL ionising luminosity, $n_e$ is the electron density of the illuminated disc, and $R$ is the distance from the ionising source to the disc. We computed the electron density corresponding to distances ranging from 3 km to 10 km for a BB luminosity associated with the NS in the 7--30 keV ionising energy range, $L_{BL}=2 \times 10^{36}$~erg~s$^{-1}$ and a degree of ionisation equal to $\sim~200$~erg~cm~s$^{-1}$. We find that the electron density at a distance of 10 km is $\sim~10^{22}$~cm$^{-3}$ and at a distance of 3 km is $\sim~10^{23}$ cm$^{-3}$. 
To have a density of 10$^{19}$~cm$^{-3}$ using the same values as in the previous case would require a distance of $\sim$ 310 km, which is inconsistent with the expected distance of the BL.

From the \cite{Shakura1973A&A....24..337S} model \citep[equation 5.49 of][]{Frank2002apa..book.....F} we can write:

\begin{equation}
\rho = 3.1 \times 10^{-8} \alpha^{-7/10} \dot{M}_{\rm 16}^{11/20} m_1^{5/8} R_{\rm 10}^{-15/8} f^{3/5} \mathrm{g~cm^{-3}},
\end{equation}

where $R_{10}$ is the outer radius of the disk in units of $10^{10}$~cm, $m_1$ is the the mass of the central object in solar masses, $\dot{M}_{16}$ is the accretion rate in units of $10^{16}$~g~s$^{-1}$, and $f=[1-(R_{\star} / R)^{1/2}]^{1/4}$. We assume that the outer radius of the accretion disk is 10$^{10}$~cm and the size of the compact object is $1.5 \times 10^6$ cm (small deviations of the sizes around these values do not affect the result significantly). Assuming $\mu = 0.615$ (fully ionised gas), for $\alpha = 0.1$, $\alpha = 0.01$ and $\alpha = 0.001$, the electron densities are $\sim 5 \times 10^{21}$ cm$^{-3}$, $\sim 2 \times 10^{22}$ cm$^{-3}$ and $\sim 10^{23}$ cm$^{-3}$, respectively.

The high density of the accretion disc in GX 13+1 is consistent with findings from other studies on density in similar sources \citep[see, for example,][]{Tomsick2018ApJ...855....3T, Jiang2019MNRAS.484.1972J, Mondal2019MNRAS.487.5441M, Chakraborty2021MNRAS.508..475C, Connors2021ApJ...909..146C}. 

There are two essential aspects of the electron density parameter associated with the {\tt relxillNS} model that require emphasis. Firstly, for the combination of parameters found in our best-fitting models, the electron density has a minor impact on energies above 3 keV (this effect is discernible only through a slight flux re-scaling in the spectrum) whereas, at energies below 3 keV, the impact of the electron density parameter is significant. Hence, to disentangle this issue, it would be crucial to study this type of sources in a softer band than that of {\em NuSTAR}, using complementary instruments such as {\sl NICER}. Secondly, given the current limitations in the atomic data available in {\tt XSTAR} \citep{Bautista2001ApJS..134..139B, Mendoza2021Atoms...9...12M}, which is used to compute the reflection tables, the electron density in the {\tt relxillNS} model extends only up to 10$^{19}$~cm$^{-3}$, which is probably 2-3 orders of magnitude below the actual density in such systems.

Previous studies by \citet{Day1991MNRAS.253P..35D}, \citet{Brandt1994MNRAS.268.1051B} and \citet{Popham2001ApJ...547..355P} have hypothesised that the BL may be the driving force behind the emission of Fe~K radiation in the inner accretion disc. Our findings support such a hypothesis. However, it is also possible that the broad Fe line could be caused by a disc wind, as proposed by \citet{Diaztrigo2012A&A...543A..50D}. In light of our results, we assert that the most probable source of the broadened Fe line is the reflection off the accretion disc, in light of the resulting relativistic effects. The presence of a Compton hump in the energy range of 10--25 keV further reinforces this conclusion.

\section{Conclusions} \label{sec:Conclusions}

The following points summarise the key results of this paper:

 \begin{enumerate}
 
\item Our analysis of the colour-colour diagram of the source indicates that GX~13+1 underwent a typical transition from the normal to the flaring branch during the observation.
\item Our findings highlight the importance of being cautious when selecting between emission or absorption lines in mid-resolution spectra, as both options can produce statistically equal satisfactory fits that have the potential to impact the continuum and other spectral components. This issue is particularly relevant for newly discovered sources for which there is no independent constraint of the system inclination, and underscores the need for careful evaluation when interpreting the spectra.
\item The inner disc radius is $R_{\rm in} \lesssim 9.6~r_g$ for both branches, which allowed us to constrain the magnetic field strength to $B \lesssim 1.8 \times 10^8$~G.
\item Evidence of a hot wind was found through photo-ionised absorption of Fe and Ni. The Ni/Fe overabundance was found to be $\sim$6 times solar.
\item The distance between the ionising source and the slab of ionised absorbing material was inferred to $\sim 4-40 \times 10^5$~km.
\item We constrained the boundary layer width to $\sim$~3~km, and the NS radius to $R_{\rm NS}\lesssim 16$~km.
\item We obtained a high electron density in the accretion disc, $n_e \sim 10^{22}-10^{23}$~cm$^{-3}$, well above the densities provided by {\tt relxillNS}, consistent with values expected in a standard disc \citep{Shakura1973A&A....24..337S}.

\end{enumerate}

\section*{Acknowledgements}
We thank Renee Ludlam for her valuable comments as the reviewer, which helped to improve considerably this manuscript. EAS is a fellow of the Consejo Interuniversitario Nacional (CIN), Argentina. JAC and FG are CONICET researchers. FAF is a fellow of CONICET. FAF, JAC, and FG acknowledge support by PIP 0113 (CONICET) and PICT-2017-2865 (ANPCyT). FG was also supported by PIBAA 1275 (CONICET). MM acknowledges the research programme Athena with project number 184.034.002, which is (partly) financed by the Dutch Research Council (NWO). JAC is a Mar\'ia Zambrano researcher fellow funded by the European Union -NextGenerationEU- (UJAR02MZ). JAC, JM, PLLE and FG also acknowledge support by grant PID2019-105510GB-C32/AEI/10.13039/501100011033 from the Agencia Estatal de Investigaci\'on of the Spanish Ministerio de Ciencia, Innovaci\'on y Universidades, and by Consejer\'{\i}a de Econom\'{\i}a, Innovaci\'on, Ciencia y Empleo of Junta de Andaluc\'{\i}a as research group FQM-322, as well as FEDER funds.  

\section*{Data Availability}

This research has made use of data obtained from the High Energy Astrophysics Science Archive Research Center (HEASARC),
provided by NASA’s Goddard Space Flight Center.


\bibliographystyle{mnras}
\bibliography{biblio}

\begin{thebibliography}{}
\makeatletter
\relax
\def\mn@urlcharsother{\let\do\@makeother \do\$\do\&\do\#\do\^\do\_\do\%\do\~}
\def\mn@doi{\begingroup\mn@urlcharsother \@ifnextchar [ {\mn@doi@}
  {\mn@doi@[]}}
\def\mn@doi@[#1]#2{\def\@tempa{#1}\ifx\@tempa\@empty \href
  {http://dx.doi.org/#2} {doi:#2}\else \href {http://dx.doi.org/#2} {#1}\fi
  \endgroup}
\def\mn@eprint#1#2{\mn@eprint@#1:#2::\@nil}
\def\mn@eprint@arXiv#1{\href {http://arxiv.org/abs/#1} {{\tt arXiv:#1}}}
\def\mn@eprint@dblp#1{\href {http://dblp.uni-trier.de/rec/bibtex/#1.xml}
  {dblp:#1}}
\def\mn@eprint@#1:#2:#3:#4\@nil{\def\@tempa {#1}\def\@tempb {#2}\def\@tempc
  {#3}\ifx \@tempc \@empty \let \@tempc \@tempb \let \@tempb \@tempa \fi \ifx
  \@tempb \@empty \def\@tempb {arXiv}\fi \@ifundefined
  {mn@eprint@\@tempb}{\@tempb:\@tempc}{\expandafter \expandafter \csname
  mn@eprint@\@tempb\endcsname \expandafter{\@tempc}}}

\bibitem[\protect\citeauthoryear{{Agrawal} \& {Misra}}{{Agrawal} \&
  {Misra}}{2009}]{Agrawal2009MNRAS.398.1352A}
{Agrawal} V.~K.,  {Misra} R.,  2009, \mn@doi [\mnras]
  {10.1111/j.1365-2966.2009.15014.x}, \href
  {https://ui.adsabs.harvard.edu/abs/2009MNRAS.398.1352A} {398, 1352}

\bibitem[\protect\citeauthoryear{{Agrawal} \& {Nandi}}{{Agrawal} \&
  {Nandi}}{2020}]{Agrawal2020MNRAS.497.3726A}
{Agrawal} V.~K.,  {Nandi} A.,  2020, \mn@doi [\mnras] {10.1093/mnras/staa2063},
  \href {https://ui.adsabs.harvard.edu/abs/2020MNRAS.497.3726A} {497, 3726}

\bibitem[\protect\citeauthoryear{{Agrawal} \& {Sreekumar}}{{Agrawal} \&
  {Sreekumar}}{2003a}]{Agrawal2003MNRAS.346..933A}
{Agrawal} V.~K.,  {Sreekumar} P.,  2003a, \mn@doi [\mnras]
  {10.1111/j.1365-2966.2003.07147.x}, \href
  {https://ui.adsabs.harvard.edu/abs/2003MNRAS.346..933A} {346, 933}

\bibitem[\protect\citeauthoryear{{Agrawal} \& {Sreekumar}}{{Agrawal} \&
  {Sreekumar}}{2003b}]{2003MNRAS.346..933A}
{Agrawal} V.~K.,  {Sreekumar} P.,  2003b, \mn@doi [\mnras]
  {10.1111/j.1365-2966.2003.07147.x}, \href
  {https://ui.adsabs.harvard.edu/abs/2003MNRAS.346..933A} {346, 933}

\bibitem[\protect\citeauthoryear{{Anitra} et~al.,}{{Anitra}
  et~al.}{2021}]{Anitra2021A&A...654A.160A}
{Anitra} A.,  et~al., 2021, \mn@doi [\aap] {10.1051/0004-6361/202140557}, \href
  {https://ui.adsabs.harvard.edu/abs/2021A&A...654A.160A} {654, A160}

\bibitem[\protect\citeauthoryear{{Arnaud}}{{Arnaud}}{1996}]{1996ASPC..101...17A}
{Arnaud} K.~A.,  1996, in {Jacoby} G.~H.,  {Barnes} J.,  eds,  Astronomical
  Society of the Pacific Conference Series Vol. 101, Astronomical Data Analysis
  Software and Systems V. p.~17

\bibitem[\protect\citeauthoryear{{Bandyopadhyay}, {Charles}, {Shahbaz}  \&
  {Wagner}}{{Bandyopadhyay} et~al.}{2002}]{2002ApJ...570..793B}
{Bandyopadhyay} R.~M.,  {Charles} P.~A.,  {Shahbaz} T.,   {Wagner} R.~M.,
  2002, \mn@doi [\apj] {10.1086/339776}, \href
  {https://ui.adsabs.harvard.edu/abs/2002ApJ...570..793B} {570, 793}

\bibitem[\protect\citeauthoryear{{Barret}}{{Barret}}{2001}]{Barret2001AdSpR..28..307B}
{Barret} D.,  2001, \mn@doi [Advances in Space Research]
  {10.1016/S0273-1177(01)00414-8}, \href
  {https://ui.adsabs.harvard.edu/abs/2001AdSpR..28..307B} {28, 307}

\bibitem[\protect\citeauthoryear{{Barret}, {Olive}, {Boirin}, {Done}, {Skinner}
   \& {Grindlay}}{{Barret} et~al.}{2000}]{Barret2000ApJ...533..329B}
{Barret} D.,  {Olive} J.~F.,  {Boirin} L.,  {Done} C.,  {Skinner} G.~K.,
  {Grindlay} J.~E.,  2000, \mn@doi [\apj] {10.1086/308651}, \href
  {https://ui.adsabs.harvard.edu/abs/2000ApJ...533..329B} {533, 329}

\bibitem[\protect\citeauthoryear{{Bautista} \& {Kallman}}{{Bautista} \&
  {Kallman}}{2001}]{Bautista2001ApJS..134..139B}
{Bautista} M.~A.,  {Kallman} T.~R.,  2001, \mn@doi [\apjs] {10.1086/320363},
  \href {https://ui.adsabs.harvard.edu/abs/2001ApJS..134..139B} {134, 139}

\bibitem[\protect\citeauthoryear{{Boirin}, {M{\'e}ndez}, {D{\'\i}az Trigo},
  {Parmar}  \& {Kaastra}}{{Boirin} et~al.}{2005}]{Boirin2005A&A...436..195B}
{Boirin} L.,  {M{\'e}ndez} M.,  {D{\'\i}az Trigo} M.,  {Parmar} A.~N.,
  {Kaastra} J.~S.,  2005, \mn@doi [\aap] {10.1051/0004-6361:20041940}, \href
  {https://ui.adsabs.harvard.edu/abs/2005A&A...436..195B} {436, 195}

\bibitem[\protect\citeauthoryear{{Brandt} \& {Matt}}{{Brandt} \&
  {Matt}}{1994}]{Brandt1994MNRAS.268.1051B}
{Brandt} W.~M.,  {Matt} G.,  1994, \mn@doi [\mnras] {10.1093/mnras/268.4.1051},
  \href {https://ui.adsabs.harvard.edu/abs/1994MNRAS.268.1051B} {268, 1051}

\bibitem[\protect\citeauthoryear{{Cackett}, {Altamirano}, {Patruno}, {Miller},
  {Reynolds}, {Linares}  \& {Wijnands}}{{Cackett}
  et~al.}{2009}]{Cackett2009ApJ...694L..21C}
{Cackett} E.~M.,  {Altamirano} D.,  {Patruno} A.,  {Miller} J.~M.,  {Reynolds}
  M.,  {Linares} M.,   {Wijnands} R.,  2009, \mn@doi [\apjl]
  {10.1088/0004-637X/694/1/L21}, \href
  {https://ui.adsabs.harvard.edu/abs/2009ApJ...694L..21C} {694, L21}

\bibitem[\protect\citeauthoryear{{Cackett} et~al.,}{{Cackett}
  et~al.}{2010}]{Cackett2010ApJ...720..205C}
{Cackett} E.~M.,  et~al., 2010, \mn@doi [\apj] {10.1088/0004-637X/720/1/205},
  \href {https://ui.adsabs.harvard.edu/abs/2010ApJ...720..205C} {720, 205}

\bibitem[\protect\citeauthoryear{{Chakraborty}, {Ratheesh}, {Bhattacharyya},
  {Tomsick}, {Tombesi}, {Fukumura}  \& {Jaisawal}}{{Chakraborty}
  et~al.}{2021}]{Chakraborty2021MNRAS.508..475C}
{Chakraborty} S.,  {Ratheesh} A.,  {Bhattacharyya} S.,  {Tomsick} J.~A.,
  {Tombesi} F.,  {Fukumura} K.,   {Jaisawal} G.~K.,  2021, \mn@doi [\mnras]
  {10.1093/mnras/stab2530}, \href
  {https://ui.adsabs.harvard.edu/abs/2021MNRAS.508..475C} {508, 475}

\bibitem[\protect\citeauthoryear{{Chen}, {Zhang}  \& {Ding}}{{Chen}
  et~al.}{2006}]{Chen2006ApJ...650..299C}
{Chen} X.,  {Zhang} S.~N.,   {Ding} G.~Q.,  2006, \mn@doi [\apj]
  {10.1086/506338}, \href
  {https://ui.adsabs.harvard.edu/abs/2006ApJ...650..299C} {650, 299}

\bibitem[\protect\citeauthoryear{{Connors} et~al.,}{{Connors}
  et~al.}{2021}]{Connors2021ApJ...909..146C}
{Connors} R. M.~T.,  et~al., 2021, \mn@doi [\apj] {10.3847/1538-4357/abdd2c},
  \href {https://ui.adsabs.harvard.edu/abs/2021ApJ...909..146C} {909, 146}

\bibitem[\protect\citeauthoryear{{Corbet}, {Pearlman}, {Buxton}  \&
  {Levine}}{{Corbet} et~al.}{2010}]{2010ApJ...719..979C}
{Corbet} R. H.~D.,  {Pearlman} A.~B.,  {Buxton} M.,   {Levine} A.~M.,  2010,
  \mn@doi [\apj] {10.1088/0004-637X/719/1/979}, \href
  {https://ui.adsabs.harvard.edu/abs/2010ApJ...719..979C} {719, 979}

\bibitem[\protect\citeauthoryear{{Coughenour}, {Cackett}, {Miller}  \&
  {Ludlam}}{{Coughenour} et~al.}{2018}]{Coughenour2018ApJ...867...64C}
{Coughenour} B.~M.,  {Cackett} E.~M.,  {Miller} J.~M.,   {Ludlam} R.~M.,  2018,
  \mn@doi [\apj] {10.3847/1538-4357/aae098}, \href
  {https://ui.adsabs.harvard.edu/abs/2018ApJ...867...64C} {867, 64}

\bibitem[\protect\citeauthoryear{{D'A{\`\i}}, {Iaria}, {Di Salvo}, {Matt}  \&
  {Robba}}{{D'A{\`\i}} et~al.}{2009}]{Daia2009ApJ...693L...1D}
{D'A{\`\i}} A.,  {Iaria} R.,  {Di Salvo} T.,  {Matt} G.,   {Robba} N.~R.,
  2009, \mn@doi [\apjl] {10.1088/0004-637X/693/1/L1}, \href
  {https://ui.adsabs.harvard.edu/abs/2009ApJ...693L...1D} {693, L1}

\bibitem[\protect\citeauthoryear{{D'A{\`\i}}, {Iaria}, {Di Salvo}, {Riggio},
  {Burderi}  \& {Robba}}{{D'A{\`\i}} et~al.}{2014}]{2014A&A...564A..62D}
{D'A{\`\i}} A.,  {Iaria} R.,  {Di Salvo} T.,  {Riggio} A.,  {Burderi} L.,
  {Robba} N.~R.,  2014, \mn@doi [\aap] {10.1051/0004-6361/201322044}, \href
  {https://ui.adsabs.harvard.edu/abs/2014A&A...564A..62D} {564, A62}

\bibitem[\protect\citeauthoryear{{Dauser}, {Wilms}, {Reynolds}  \&
  {Brenneman}}{{Dauser} et~al.}{2010}]{Dauser2010MNRAS.409.1534D}
{Dauser} T.,  {Wilms} J.,  {Reynolds} C.~S.,   {Brenneman} L.~W.,  2010,
  \mn@doi [\mnras] {10.1111/j.1365-2966.2010.17393.x}, \href
  {https://ui.adsabs.harvard.edu/abs/2010MNRAS.409.1534D} {409, 1534}

\bibitem[\protect\citeauthoryear{{Dauser}, {Garcia}, {Parker}, {Fabian}  \&
  {Wilms}}{{Dauser} et~al.}{2014}]{Dauser2014MNRAS.444L.100D}
{Dauser} T.,  {Garcia} J.,  {Parker} M.~L.,  {Fabian} A.~C.,   {Wilms} J.,
  2014, \mn@doi [\mnras] {10.1093/mnrasl/slu125}, \href
  {https://ui.adsabs.harvard.edu/abs/2014MNRAS.444L.100D} {444, L100}

\bibitem[\protect\citeauthoryear{{Dauser}, {Garc{\'\i}a}, {Walton}, {Eikmann},
  {Kallman}, {McClintock}  \& {Wilms}}{{Dauser}
  et~al.}{2016}]{Dauser2016A&A...590A..76D}
{Dauser} T.,  {Garc{\'\i}a} J.,  {Walton} D.~J.,  {Eikmann} W.,  {Kallman} T.,
  {McClintock} J.,   {Wilms} J.,  2016, \mn@doi [\aap]
  {10.1051/0004-6361/201628135}, \href
  {https://ui.adsabs.harvard.edu/abs/2016A&A...590A..76D} {590, A76}

\bibitem[\protect\citeauthoryear{{Day} \& {Done}}{{Day} \&
  {Done}}{1991}]{Day1991MNRAS.253P..35D}
{Day} C.~S.~R.,  {Done} C.,  1991, \mn@doi [\mnras] {10.1093/mnras/253.1.35P},
  \href {https://ui.adsabs.harvard.edu/abs/1991MNRAS.253P..35D} {253, 35P}

\bibitem[\protect\citeauthoryear{{Degenaar}, {Miller}, {Chakrabarty},
  {Harrison}, {Kara}  \& {Fabian}}{{Degenaar}
  et~al.}{2015}]{Degenaar2015MNRAS.451L..85D}
{Degenaar} N.,  {Miller} J.~M.,  {Chakrabarty} D.,  {Harrison} F.~A.,  {Kara}
  E.,   {Fabian} A.~C.,  2015, \mn@doi [\mnras] {10.1093/mnrasl/slv072}, \href
  {https://ui.adsabs.harvard.edu/abs/2015MNRAS.451L..85D} {451, L85}

\bibitem[\protect\citeauthoryear{{Degenaar} et~al.,}{{Degenaar}
  et~al.}{2016}]{Degenaar2016MNRAS.461.4049D}
{Degenaar} N.,  et~al., 2016, \mn@doi [\mnras] {10.1093/mnras/stw1593}, \href
  {https://ui.adsabs.harvard.edu/abs/2016MNRAS.461.4049D} {461, 4049}

\bibitem[\protect\citeauthoryear{{Di Salvo} \& {Burderi}}{{Di Salvo} \&
  {Burderi}}{2003}]{Disalvo2003A&A...397..723D}
{Di Salvo} T.,  {Burderi} L.,  2003, \mn@doi [\aap]
  {10.1051/0004-6361:20021491}, \href
  {https://ui.adsabs.harvard.edu/abs/2003A&A...397..723D} {397, 723}

\bibitem[\protect\citeauthoryear{{Di Salvo} \& {Sanna}}{{Di Salvo} \&
  {Sanna}}{2022}]{DiSalvo2022ASSL..465...87D}
{Di Salvo} T.,  {Sanna} A.,  2022, in {Bhattacharyya} S.,  {Papitto} A.,
  {Bhattacharya} D.,  eds,  Astrophysics and Space Science Library Vol. 465,
  Astrophysics and Space Science Library. pp 87--124,
  \mn@doi{10.1007/978-3-030-85198-9_4}

\bibitem[\protect\citeauthoryear{{Di Salvo} et~al.,}{{Di Salvo}
  et~al.}{2000}]{DiSalvo2000ApJ...544L.119D}
{Di Salvo} T.,  et~al., 2000, \mn@doi [\apjl] {10.1086/317309}, \href
  {https://ui.adsabs.harvard.edu/abs/2000ApJ...544L.119D} {544, L119}

\bibitem[\protect\citeauthoryear{{Di Salvo}, {Robba}, {Iaria}, {Stella},
  {Burderi}  \& {Israel}}{{Di Salvo} et~al.}{2001}]{Disalvo2001ApJ...554...49D}
{Di Salvo} T.,  {Robba} N.~R.,  {Iaria} R.,  {Stella} L.,  {Burderi} L.,
  {Israel} G.~L.,  2001, \mn@doi [\apj] {10.1086/321353}, \href
  {https://ui.adsabs.harvard.edu/abs/2001ApJ...554...49D} {554, 49}

\bibitem[\protect\citeauthoryear{{Di Salvo} et~al.,}{{Di Salvo}
  et~al.}{2002}]{Disalvo2002A&A...386..535D}
{Di Salvo} T.,  et~al., 2002, \mn@doi [\aap] {10.1051/0004-6361:20020238},
  \href {https://ui.adsabs.harvard.edu/abs/2002A&A...386..535D} {386, 535}

\bibitem[\protect\citeauthoryear{{D{\'\i}az Trigo}, {Parmar}, {Boirin},
  {M{\'e}ndez}  \& {Kaastra}}{{D{\'\i}az Trigo}
  et~al.}{2006}]{DiazTrigo2006A&A...445..179D}
{D{\'\i}az Trigo} M.,  {Parmar} A.~N.,  {Boirin} L.,  {M{\'e}ndez} M.,
  {Kaastra} J.~S.,  2006, \mn@doi [\aap] {10.1051/0004-6361:20053586}, \href
  {https://ui.adsabs.harvard.edu/abs/2006A&A...445..179D} {445, 179}

\bibitem[\protect\citeauthoryear{{D{\'\i}az Trigo}, {Sidoli}, {Boirin}  \&
  {Parmar}}{{D{\'\i}az Trigo} et~al.}{2012a}]{2012A&A...543A..50D}
{D{\'\i}az Trigo} M.,  {Sidoli} L.,  {Boirin} L.,   {Parmar} A.~N.,  2012a,
  \mn@doi [\aap] {10.1051/0004-6361/201219049}, \href
  {https://ui.adsabs.harvard.edu/abs/2012A&A...543A..50D} {543, A50}

\bibitem[\protect\citeauthoryear{{D{\'\i}az Trigo}, {Sidoli}, {Boirin}  \&
  {Parmar}}{{D{\'\i}az Trigo} et~al.}{2012b}]{Diaztrigo2012A&A...543A..50D}
{D{\'\i}az Trigo} M.,  {Sidoli} L.,  {Boirin} L.,   {Parmar} A.~N.,  2012b,
  \mn@doi [\aap] {10.1051/0004-6361/201219049}, \href
  {https://ui.adsabs.harvard.edu/abs/2012A&A...543A..50D} {543, A50}

\bibitem[\protect\citeauthoryear{{Ding} \& {Huang}}{{Ding} \&
  {Huang}}{2015}]{Ding2015JApA...36..335D}
{Ding} G.~Q.,  {Huang} C.~P.,  2015, \mn@doi [Journal of Astrophysics and
  Astronomy] {10.1007/s12036-015-9340-2}, \href
  {https://ui.adsabs.harvard.edu/abs/2015JApA...36..335D} {36, 335}

\bibitem[\protect\citeauthoryear{{Ding}, {Zhang}, {Wang}, {Qu}  \&
  {Yan}}{{Ding} et~al.}{2011}]{Ding2011AJ....142...34D}
{Ding} G.~Q.,  {Zhang} S.~N.,  {Wang} N.,  {Qu} J.~L.,   {Yan} S.~P.,  2011,
  \mn@doi [\aj] {10.1088/0004-6256/142/2/34}, \href
  {https://ui.adsabs.harvard.edu/abs/2011AJ....142...34D} {142, 34}

\bibitem[\protect\citeauthoryear{{Fabian}, {Rees}, {Stella}  \&
  {White}}{{Fabian} et~al.}{1989a}]{1989MNRAS.238..729F}
{Fabian} A.~C.,  {Rees} M.~J.,  {Stella} L.,   {White} N.~E.,  1989a, \mn@doi
  [\mnras] {10.1093/mnras/238.3.729}, \href
  {https://ui.adsabs.harvard.edu/abs/1989MNRAS.238..729F} {238, 729}

\bibitem[\protect\citeauthoryear{{Fabian}, {Rees}, {Stella}  \&
  {White}}{{Fabian} et~al.}{1989b}]{Fabian1989MNRAS.238..729F}
{Fabian} A.~C.,  {Rees} M.~J.,  {Stella} L.,   {White} N.~E.,  1989b, \mn@doi
  [\mnras] {10.1093/mnras/238.3.729}, \href
  {https://ui.adsabs.harvard.edu/abs/1989MNRAS.238..729F} {238, 729}

\bibitem[\protect\citeauthoryear{{Fleischman}}{{Fleischman}}{1985}]{Fleischman1985A&A...153..106F}
{Fleischman} J.~R.,  1985, \aap, \href
  {https://ui.adsabs.harvard.edu/abs/1985A&A...153..106F} {153, 106}

\bibitem[\protect\citeauthoryear{{Focke}}{{Focke}}{1996}]{focke1996ApJ...470L.127F}
{Focke} W.~B.,  1996, \mn@doi [\apjl] {10.1086/310314}, \href
  {https://ui.adsabs.harvard.edu/abs/1996ApJ...470L.127F} {470, L127}

\bibitem[\protect\citeauthoryear{{Fogantini}, {Garc{\'\i}a}, {Combi}, {Chaty},
  {Mart{\'\i}}  \& {Luque Escamilla}}{{Fogantini}
  et~al.}{2022}]{Fogantini2022arXiv221009390F}
{Fogantini} F.~A.,  {Garc{\'\i}a} F.,  {Combi} J.~A.,  {Chaty} S.,
  {Mart{\'\i}} J.,   {Luque Escamilla} P.~L.,  2022, arXiv e-prints, \href
  {https://ui.adsabs.harvard.edu/abs/2022arXiv221009390F} {p. arXiv:2210.09390}

\bibitem[\protect\citeauthoryear{Frank, King  \& Raine}{Frank
  et~al.}{2002}]{Frank2002apa..book.....F}
Frank J.,  King A.,   Raine D.,  2002, Accretion Power in Astrophysics, 3 edn.
Cambridge University Press, \mn@doi{10.1017/CBO9781139164245}

\bibitem[\protect\citeauthoryear{Fridriksson, Homan  \& Remillard}{Fridriksson
  et~al.}{2015}]{Fridriksson}
Fridriksson J.~K.,  Homan J.,   Remillard R.~A.,  2015, \mn@doi [The
  Astrophysical Journal] {10.1088/0004-637x/809/1/52}, 809, 52

\bibitem[\protect\citeauthoryear{{Galloway}, {Muno}, {Hartman}, {Psaltis}  \&
  {Chakrabarty}}{{Galloway} et~al.}{2008}]{Galloway2008ApJS..179..360G}
{Galloway} D.~K.,  {Muno} M.~P.,  {Hartman} J.~M.,  {Psaltis} D.,
  {Chakrabarty} D.,  2008, \mn@doi [\apjs] {10.1086/592044}, \href
  {https://ui.adsabs.harvard.edu/abs/2008ApJS..179..360G} {179, 360}

\bibitem[\protect\citeauthoryear{{Garc{\'\i}a}, {Dauser}, {Reynolds},
  {Kallman}, {McClintock}, {Wilms}  \& {Eikmann}}{{Garc{\'\i}a}
  et~al.}{2013}]{Garcia2013ApJ...768..146G}
{Garc{\'\i}a} J.,  {Dauser} T.,  {Reynolds} C.~S.,  {Kallman} T.~R.,
  {McClintock} J.~E.,  {Wilms} J.,   {Eikmann} W.,  2013, \mn@doi [\apj]
  {10.1088/0004-637X/768/2/146}, \href
  {https://ui.adsabs.harvard.edu/abs/2013ApJ...768..146G} {768, 146}

\bibitem[\protect\citeauthoryear{{Garc{\'\i}a}, {Dauser}, {Ludlam}, {Parker},
  {Fabian}, {Harrison}  \& {Wilms}}{{Garc{\'\i}a}
  et~al.}{2022}]{Garcia2022ApJ...926...13G}
{Garc{\'\i}a} J.~A.,  {Dauser} T.,  {Ludlam} R.,  {Parker} M.,  {Fabian} A.,
  {Harrison} F.~A.,   {Wilms} J.,  2022, \mn@doi [\apj]
  {10.3847/1538-4357/ac3cb7}, \href
  {https://ui.adsabs.harvard.edu/abs/2022ApJ...926...13G} {926, 13}

\bibitem[\protect\citeauthoryear{{George} \& {Fabian}}{{George} \&
  {Fabian}}{1991}]{George1991MNRAS.249..352G}
{George} I.~M.,  {Fabian} A.~C.,  1991, \mn@doi [\mnras]
  {10.1093/mnras/249.2.352}, \href
  {https://ui.adsabs.harvard.edu/abs/1991MNRAS.249..352G} {249, 352}

\bibitem[\protect\citeauthoryear{{Gierli{\'n}ski} \& {Done}}{{Gierli{\'n}ski}
  \& {Done}}{2002}]{Gierlinski2002MNRAS.331L..47G}
{Gierli{\'n}ski} M.,  {Done} C.,  2002, \mn@doi [\mnras]
  {10.1046/j.1365-8711.2002.05430.x}, \href
  {https://ui.adsabs.harvard.edu/abs/2002MNRAS.331L..47G} {331, L47}

\bibitem[\protect\citeauthoryear{{Gilfanov}, {Revnivtsev}  \&
  {Molkov}}{{Gilfanov} et~al.}{2003}]{Gilfanov2003A&A...410..217G}
{Gilfanov} M.,  {Revnivtsev} M.,   {Molkov} S.,  2003, \mn@doi [\aap]
  {10.1051/0004-6361:20031141}, \href
  {https://ui.adsabs.harvard.edu/abs/2003A&A...410..217G} {410, 217}

\bibitem[\protect\citeauthoryear{{Grebenev} \& {Sunyaev}}{{Grebenev} \&
  {Sunyaev}}{2002}]{Grebenev2002AstL...28..150G}
{Grebenev} S.~A.,  {Sunyaev} R.~A.,  2002, \mn@doi [Astronomy Letters]
  {10.1134/1.1458344}, \href
  {https://ui.adsabs.harvard.edu/abs/2002AstL...28..150G} {28, 150}

\bibitem[\protect\citeauthoryear{{Guilbert} \& {Rees}}{{Guilbert} \&
  {Rees}}{1988}]{1988MNRAS.233..475G}
{Guilbert} P.~W.,  {Rees} M.~J.,  1988, \mn@doi [\mnras]
  {10.1093/mnras/233.2.475}, \href
  {https://ui.adsabs.harvard.edu/abs/1988MNRAS.233..475G} {233, 475}

\bibitem[\protect\citeauthoryear{{Harrison} et~al.,}{{Harrison}
  et~al.}{2013}]{2013ApJ...770..103H}
{Harrison} F.~A.,  et~al., 2013, \mn@doi [\apj] {10.1088/0004-637X/770/2/103},
  \href {https://ui.adsabs.harvard.edu/abs/2013ApJ...770..103H} {770, 103}

\bibitem[\protect\citeauthoryear{{Hasinger} \& {van der Klis}}{{Hasinger} \&
  {van der Klis}}{1989}]{1989A&A...225...79H}
{Hasinger} G.,  {van der Klis} M.,  1989, \aap, \href
  {https://ui.adsabs.harvard.edu/abs/1989A&A...225...79H} {225, 79}

\bibitem[\protect\citeauthoryear{{Hiemstra}, {M{\'e}ndez}, {Done}, {D{\'\i}az
  Trigo}, {Altamirano}  \& {Casella}}{{Hiemstra}
  et~al.}{2011}]{Hiemstra2011MNRAS.411..137H}
{Hiemstra} B.,  {M{\'e}ndez} M.,  {Done} C.,  {D{\'\i}az Trigo} M.,
  {Altamirano} D.,   {Casella} P.,  2011, \mn@doi [\mnras]
  {10.1111/j.1365-2966.2010.17661.x}, \href
  {https://ui.adsabs.harvard.edu/abs/2011MNRAS.411..137H} {411, 137}

\bibitem[\protect\citeauthoryear{{Homan}, {van der Klis}, {Jonker}, {Wijnands},
  {Kuulkers}, {M{\'e}ndez}  \& {Lewin}}{{Homan}
  et~al.}{2002}]{2002ApJ...568..878H}
{Homan} J.,  {van der Klis} M.,  {Jonker} P.~G.,  {Wijnands} R.,  {Kuulkers}
  E.,  {M{\'e}ndez} M.,   {Lewin} W. H.~G.,  2002, \mn@doi [\apj]
  {10.1086/339057}, \href
  {https://ui.adsabs.harvard.edu/abs/2002ApJ...568..878H} {568, 878}

\bibitem[\protect\citeauthoryear{{Homan}, {Wijnands}, {Rupen}, {Fender},
  {Hjellming}, {di Salvo}  \& {van der Klis}}{{Homan}
  et~al.}{2004}]{2004A&A...418..255H}
{Homan} J.,  {Wijnands} R.,  {Rupen} M.~P.,  {Fender} R.,  {Hjellming} R.~M.,
  {di Salvo} T.,   {van der Klis} M.,  2004, \mn@doi [\aap]
  {10.1051/0004-6361:20034258}, \href
  {https://ui.adsabs.harvard.edu/abs/2004A&A...418..255H} {418, 255}

\bibitem[\protect\citeauthoryear{{Homan}, {Steiner}, {Lin}, {Fridriksson},
  {Remillard}, {Miller}  \& {Ludlam}}{{Homan}
  et~al.}{2018}]{Homan2018ApJ...853..157H}
{Homan} J.,  {Steiner} J.~F.,  {Lin} D.,  {Fridriksson} J.~K.,  {Remillard}
  R.~A.,  {Miller} J.~M.,   {Ludlam} R.~M.,  2018, \mn@doi [\apj]
  {10.3847/1538-4357/aaa439}, \href
  {https://ui.adsabs.harvard.edu/abs/2018ApJ...853..157H} {853, 157}

\bibitem[\protect\citeauthoryear{{Huppenkothen} et~al.,}{{Huppenkothen}
  et~al.}{2019}]{stingray}
{Huppenkothen} D.,  et~al., 2019, \mn@doi [\apj] {10.3847/1538-4357/ab258d},
  \href {https://ui.adsabs.harvard.edu/abs/2019ApJ...881...39H} {881, 39}

\bibitem[\protect\citeauthoryear{{Iaria}, {D'A{\'\i}}, {di Salvo}, {Robba},
  {Riggio}, {Papitto}  \& {Burderi}}{{Iaria}
  et~al.}{2009}]{Iaria2009A&A...505.1143I}
{Iaria} R.,  {D'A{\'\i}} A.,  {di Salvo} T.,  {Robba} N.~R.,  {Riggio} A.,
  {Papitto} A.,   {Burderi} L.,  2009, \mn@doi [\aap]
  {10.1051/0004-6361/200911936}, \href
  {https://ui.adsabs.harvard.edu/abs/2009A&A...505.1143I} {505, 1143}

\bibitem[\protect\citeauthoryear{{Iaria}, {Di Salvo}, {Burderi}, {Riggio},
  {D'A{\`\i}}  \& {Robba}}{{Iaria} et~al.}{2014}]{2014A&A...561A..99I}
{Iaria} R.,  {Di Salvo} T.,  {Burderi} L.,  {Riggio} A.,  {D'A{\`\i}} A.,
  {Robba} N.~R.,  2014, \mn@doi [\aap] {10.1051/0004-6361/201322328}, \href
  {https://ui.adsabs.harvard.edu/abs/2014A&A...561A..99I} {561, A99}

\bibitem[\protect\citeauthoryear{{Ibragimov} \& {Poutanen}}{{Ibragimov} \&
  {Poutanen}}{2009}]{Ibragimov2009MNRAS.400..492I}
{Ibragimov} A.,  {Poutanen} J.,  2009, \mn@doi [\mnras]
  {10.1111/j.1365-2966.2009.15477.x}, \href
  {https://ui.adsabs.harvard.edu/abs/2009MNRAS.400..492I} {400, 492}

\bibitem[\protect\citeauthoryear{{Inogamov} \& {Sunyaev}}{{Inogamov} \&
  {Sunyaev}}{1999}]{Inogamov1999AstL...25..269I}
{Inogamov} N.~A.,  {Sunyaev} R.~A.,  1999, Astronomy Letters, \href
  {https://ui.adsabs.harvard.edu/abs/1999AstL...25..269I} {25, 269}

\bibitem[\protect\citeauthoryear{{Jackson}, {Church}  \&
  {Ba{\l}uci{\'n}ska-Church}}{{Jackson}
  et~al.}{2009}]{Jackson2009A&A...494.1059J}
{Jackson} N.~K.,  {Church} M.~J.,   {Ba{\l}uci{\'n}ska-Church} M.,  2009,
  \mn@doi [\aap] {10.1051/0004-6361:20079234}, \href
  {https://ui.adsabs.harvard.edu/abs/2009A&A...494.1059J} {494, 1059}

\bibitem[\protect\citeauthoryear{{Jiang}, {Fabian}, {Wang}, {Walton},
  {Garc{\'\i}a}, {Parker}, {Steiner}  \& {Tomsick}}{{Jiang}
  et~al.}{2019}]{Jiang2019MNRAS.484.1972J}
{Jiang} J.,  {Fabian} A.~C.,  {Wang} J.,  {Walton} D.~J.,  {Garc{\'\i}a} J.~A.,
   {Parker} M.~L.,  {Steiner} J.~F.,   {Tomsick} J.~A.,  2019, \mn@doi [\mnras]
  {10.1093/mnras/stz095}, \href
  {https://ui.adsabs.harvard.edu/abs/2019MNRAS.484.1972J} {484, 1972}

\bibitem[\protect\citeauthoryear{{Kallman} \& {Bautista}}{{Kallman} \&
  {Bautista}}{2001}]{Kallman2001ApJS..133..221K}
{Kallman} T.,  {Bautista} M.,  2001, \mn@doi [\apjs] {10.1086/319184}, \href
  {https://ui.adsabs.harvard.edu/abs/2001ApJS..133..221K} {133, 221}

\bibitem[\protect\citeauthoryear{{Kallman}, {Liedahl}, {Osterheld}, {Goldstein}
   \& {Kahn}}{{Kallman} et~al.}{1996}]{Kallman1996ApJ...465..994K}
{Kallman} T.~R.,  {Liedahl} D.,  {Osterheld} A.,  {Goldstein} W.,   {Kahn} S.,
  1996, \mn@doi [\apj] {10.1086/177485}, \href
  {https://ui.adsabs.harvard.edu/abs/1996ApJ...465..994K} {465, 994}

\bibitem[\protect\citeauthoryear{{Kallman}, {Palmeri}, {Bautista}, {Mendoza}
  \& {Krolik}}{{Kallman} et~al.}{2004}]{Kallamn2004ApJS..155..675K}
{Kallman} T.~R.,  {Palmeri} P.,  {Bautista} M.~A.,  {Mendoza} C.,   {Krolik}
  J.~H.,  2004, \mn@doi [\apjs] {10.1086/424039}, \href
  {https://ui.adsabs.harvard.edu/abs/2004ApJS..155..675K} {155, 675}

\bibitem[\protect\citeauthoryear{{Kallman}, {Bautista}, {Goriely}, {Mendoza},
  {Miller}, {Palmeri}, {Quinet}  \& {Raymond}}{{Kallman}
  et~al.}{2009}]{Kallman2009ApJ...701..865K}
{Kallman} T.~R.,  {Bautista} M.~A.,  {Goriely} S.,  {Mendoza} C.,  {Miller}
  J.~M.,  {Palmeri} P.,  {Quinet} P.,   {Raymond} J.,  2009, \mn@doi [\apj]
  {10.1088/0004-637X/701/2/865}, \href
  {https://ui.adsabs.harvard.edu/abs/2009ApJ...701..865K} {701, 865}

\bibitem[\protect\citeauthoryear{{Koljonen} \& {Tomsick}}{{Koljonen} \&
  {Tomsick}}{2020}]{Koljonen2020A&A...639A..13K}
{Koljonen} K.~I.~I.,  {Tomsick} J.~A.,  2020, \mn@doi [\aap]
  {10.1051/0004-6361/202037882}, \href
  {https://ui.adsabs.harvard.edu/abs/2020A&A...639A..13K} {639, A13}

\bibitem[\protect\citeauthoryear{{Kuulkers}, {den Hartog}, {in't Zand},
  {Verbunt}, {Harris}  \& {Cocchi}}{{Kuulkers}
  et~al.}{2003}]{Kuulkers2003A&A...399..663K}
{Kuulkers} E.,  {den Hartog} P.~R.,  {in't Zand} J.~J.~M.,  {Verbunt} F.~W.~M.,
   {Harris} W.~E.,   {Cocchi} M.,  2003, \mn@doi [\aap]
  {10.1051/0004-6361:20021781}, \href
  {https://ui.adsabs.harvard.edu/abs/2003A&A...399..663K} {399, 663}

\bibitem[\protect\citeauthoryear{{Lamb}, {Pethick}  \& {Pines}}{{Lamb}
  et~al.}{1973}]{Lamb1973ApJ...184..271L}
{Lamb} F.~K.,  {Pethick} C.~J.,   {Pines} D.,  1973, \mn@doi [\apj]
  {10.1086/152325}, \href
  {https://ui.adsabs.harvard.edu/abs/1973ApJ...184..271L} {184, 271}

\bibitem[\protect\citeauthoryear{{Lavagetto}, {Iaria}, {D'A{\i}}, {di Salvo}
  \& {Robba}}{{Lavagetto} et~al.}{2008}]{Lavagetto2008A&A...478..181L}
{Lavagetto} G.,  {Iaria} R.,  {D'A{\i}} A.,  {di Salvo} T.,   {Robba} N.~R.,
  2008, \mn@doi [\aap] {10.1051/0004-6361:20078027}, \href
  {https://ui.adsabs.harvard.edu/abs/2008A&A...478..181L} {478, 181}

\bibitem[\protect\citeauthoryear{{Li}, {Zimmerman}, {Narayan}  \&
  {McClintock}}{{Li} et~al.}{2005}]{2005ApJS..157..335L}
{Li} L.-X.,  {Zimmerman} E.~R.,  {Narayan} R.,   {McClintock} J.~E.,  2005,
  \mn@doi [\apjs] {10.1086/428089}, \href
  {https://ui.adsabs.harvard.edu/abs/2005ApJS..157..335L} {157, 335}

\bibitem[\protect\citeauthoryear{{Lin}, {Altamirano}, {Homan}, {Remillard},
  {Wijnands}  \& {Belloni}}{{Lin} et~al.}{2009}]{Lin2009ApJ...699...60L}
{Lin} D.,  {Altamirano} D.,  {Homan} J.,  {Remillard} R.~A.,  {Wijnands} R.,
  {Belloni} T.,  2009, \mn@doi [\apj] {10.1088/0004-637X/699/1/60}, \href
  {https://ui.adsabs.harvard.edu/abs/2009ApJ...699...60L} {699, 60}

\bibitem[\protect\citeauthoryear{{Lin}, {Remillard}, {Homan}  \&
  {Barret}}{{Lin} et~al.}{2012}]{Lin2012ApJ...756...34L}
{Lin} D.,  {Remillard} R.~A.,  {Homan} J.,   {Barret} D.,  2012, \mn@doi [\apj]
  {10.1088/0004-637X/756/1/34}, \href
  {https://ui.adsabs.harvard.edu/abs/2012ApJ...756...34L} {756, 34}

\bibitem[\protect\citeauthoryear{{Ludlam}, {Miller}, {Degenaar}, {Sanna},
  {Cackett}, {Altamirano}  \& {King}}{{Ludlam}
  et~al.}{2017}]{Ludlam2017ApJ...847..135L}
{Ludlam} R.~M.,  {Miller} J.~M.,  {Degenaar} N.,  {Sanna} A.,  {Cackett} E.~M.,
   {Altamirano} D.,   {King} A.~L.,  2017, \mn@doi [\apj]
  {10.3847/1538-4357/aa8b1b}, \href
  {https://ui.adsabs.harvard.edu/abs/2017ApJ...847..135L} {847, 135}

\bibitem[\protect\citeauthoryear{{Ludlam} et~al.,}{{Ludlam}
  et~al.}{2018}]{Ludman2018ApJ...858L...5L}
{Ludlam} R.~M.,  et~al., 2018, \mn@doi [\apjl] {10.3847/2041-8213/aabee6},
  \href {https://ui.adsabs.harvard.edu/abs/2018ApJ...858L...5L} {858, L5}

\bibitem[\protect\citeauthoryear{{Ludlam} et~al.,}{{Ludlam}
  et~al.}{2019}]{Ludlam2019ApJ...873...99L}
{Ludlam} R.~M.,  et~al., 2019, \mn@doi [\apj] {10.3847/1538-4357/ab0414}, \href
  {https://ui.adsabs.harvard.edu/abs/2019ApJ...873...99L} {873, 99}

\bibitem[\protect\citeauthoryear{{Ludlam} et~al.,}{{Ludlam}
  et~al.}{2020}]{Ludlam2020ApJ...895...45L}
{Ludlam} R.~M.,  et~al., 2020, \mn@doi [\apj] {10.3847/1538-4357/ab89a6}, \href
  {https://ui.adsabs.harvard.edu/abs/2020ApJ...895...45L} {895, 45}

\bibitem[\protect\citeauthoryear{{Ludlam} et~al.,}{{Ludlam}
  et~al.}{2021}]{Ludlam2021ApJ...911..123L}
{Ludlam} R.~M.,  et~al., 2021, \mn@doi [\apj] {10.3847/1538-4357/abedb0}, \href
  {https://ui.adsabs.harvard.edu/abs/2021ApJ...911..123L} {911, 123}

\bibitem[\protect\citeauthoryear{{Mainardi} et~al.,}{{Mainardi}
  et~al.}{2010}]{2010A&A...512A..57M}
{Mainardi} L.~I.,  et~al., 2010, \mn@doi [\aap] {10.1051/0004-6361/200912921},
  \href {https://ui.adsabs.harvard.edu/abs/2010A&A...512A..57M} {512, A57}

\bibitem[\protect\citeauthoryear{{Maiolino}, {Laurent}, {Titarchuk},
  {Orlandini}  \& {Frontera}}{{Maiolino}
  et~al.}{2019}]{Maiolino2019A&A...625A...8M}
{Maiolino} T.,  {Laurent} P.,  {Titarchuk} L.,  {Orlandini} M.,   {Frontera}
  F.,  2019, \mn@doi [\aap] {10.1051/0004-6361/201833163}, \href
  {https://ui.adsabs.harvard.edu/abs/2019A&A...625A...8M} {625, A8}

\bibitem[\protect\citeauthoryear{{Matsuba}, {Dotani}, {Mitsuda}, {Asai},
  {Lewin}, {van Paradijs}  \& {van der Klis}}{{Matsuba}
  et~al.}{1995}]{Matsuba1995PASJ...47..575M}
{Matsuba} E.,  {Dotani} T.,  {Mitsuda} K.,  {Asai} K.,  {Lewin} W. H.~G.,  {van
  Paradijs} J.,   {van der Klis} M.,  1995, \pasj, \href
  {https://ui.adsabs.harvard.edu/abs/1995PASJ...47..575M} {47, 575}

\bibitem[\protect\citeauthoryear{{Mazzola} et~al.,}{{Mazzola}
  et~al.}{2019}]{Mazzola2019A&A...621A..89M}
{Mazzola} S.~M.,  et~al., 2019, \mn@doi [\aap] {10.1051/0004-6361/201732383},
  \href {https://ui.adsabs.harvard.edu/abs/2019A&A...621A..89M} {621, A89}

\bibitem[\protect\citeauthoryear{{Medvedev}, {Khabibullin}, {Sazonov},
  {Churazov}  \& {Tsygankov}}{{Medvedev}
  et~al.}{2018}]{Medvedev2018AstL...44..390M}
{Medvedev} P.~S.,  {Khabibullin} I.~I.,  {Sazonov} S.~Y.,  {Churazov} E.~M.,
  {Tsygankov} S.~S.,  2018, \mn@doi [Astronomy Letters]
  {10.1134/S1063773718060038}, \href
  {https://ui.adsabs.harvard.edu/abs/2018AstL...44..390M} {44, 390}

\bibitem[\protect\citeauthoryear{{M{\'e}ndez} \& {Belloni}}{{M{\'e}ndez} \&
  {Belloni}}{2021}]{2021ASSL..461..263M}
{M{\'e}ndez} M.,  {Belloni} T.~M.,  2021, in {Belloni} T.~M.,  {M{\'e}ndez} M.,
    {Zhang} C.,  eds,  Astrophysics and Space Science Library Vol. 461,
  Astrophysics and Space Science Library. pp 263--331 (\mn@eprint {arXiv}
  {2010.08291}), \mn@doi{10.1007/978-3-662-62110-3_6}

\bibitem[\protect\citeauthoryear{{Mendoza} et~al.,}{{Mendoza}
  et~al.}{2021}]{Mendoza2021Atoms...9...12M}
{Mendoza} C.,  et~al., 2021, \mn@doi [Atoms] {10.3390/atoms9010012}, \href
  {https://ui.adsabs.harvard.edu/abs/2021Atoms...9...12M} {9, 12}

\bibitem[\protect\citeauthoryear{{Miller}}{{Miller}}{2007}]{Miller2007ARA&A..45..441M}
{Miller} J.~M.,  2007, \mn@doi [\araa]
  {10.1146/annurev.astro.45.051806.110555}, \href
  {https://ui.adsabs.harvard.edu/abs/2007ARA&A..45..441M} {45, 441}

\bibitem[\protect\citeauthoryear{{Miller}, {Maitra}, {Cackett}, {Bhattacharyya}
   \& {Strohmayer}}{{Miller} et~al.}{2011}]{Miller2011ApJ...731L...7M}
{Miller} J.~M.,  {Maitra} D.,  {Cackett} E.~M.,  {Bhattacharyya} S.,
  {Strohmayer} T.~E.,  2011, \mn@doi [\apjl] {10.1088/2041-8205/731/1/L7},
  \href {https://ui.adsabs.harvard.edu/abs/2011ApJ...731L...7M} {731, L7}

\bibitem[\protect\citeauthoryear{{Miller} et~al.,}{{Miller}
  et~al.}{2013a}]{Miller}
{Miller} J.~M.,  et~al., 2013a, \mn@doi [\apjl] {10.1088/2041-8205/779/1/L2},
  \href {https://ui.adsabs.harvard.edu/abs/2013ApJ...779L...2M} {779, L2}

\bibitem[\protect\citeauthoryear{{Miller} et~al.,}{{Miller}
  et~al.}{2013b}]{Miller2013ApJ...779L...2M}
{Miller} J.~M.,  et~al., 2013b, \mn@doi [\apjl] {10.1088/2041-8205/779/1/L2},
  \href {https://ui.adsabs.harvard.edu/abs/2013ApJ...779L...2M} {779, L2}

\bibitem[\protect\citeauthoryear{{Mitsuda} et~al.,}{{Mitsuda}
  et~al.}{1984}]{Mitsuda1984PASJ...36..741M}
{Mitsuda} K.,  et~al., 1984, \pasj, \href
  {https://ui.adsabs.harvard.edu/abs/1984PASJ...36..741M} {36, 741}

\bibitem[\protect\citeauthoryear{{Mondal}, {Dewangan}, {Pahari}  \&
  {Raychaudhuri}}{{Mondal} et~al.}{2018}]{Mondal2018MNRAS.474.2064M}
{Mondal} A.~S.,  {Dewangan} G.~C.,  {Pahari} M.,   {Raychaudhuri} B.,  2018,
  \mn@doi [\mnras] {10.1093/mnras/stx2931}, \href
  {https://ui.adsabs.harvard.edu/abs/2018MNRAS.474.2064M} {474, 2064}

\bibitem[\protect\citeauthoryear{{Mondal}, {Dewangan}  \&
  {Raychaudhuri}}{{Mondal} et~al.}{2019}]{Mondal2019MNRAS.487.5441M}
{Mondal} A.~S.,  {Dewangan} G.~C.,   {Raychaudhuri} B.,  2019, \mn@doi [\mnras]
  {10.1093/mnras/stz1658}, \href
  {https://ui.adsabs.harvard.edu/abs/2019MNRAS.487.5441M} {487, 5441}

\bibitem[\protect\citeauthoryear{{Mondal}, {Dewangan}  \&
  {Raychaudhuri}}{{Mondal} et~al.}{2020}]{Mondal2020MNRAS.494.3177M}
{Mondal} A.~S.,  {Dewangan} G.~C.,   {Raychaudhuri} B.,  2020, \mn@doi [\mnras]
  {10.1093/mnras/staa1001}, \href
  {https://ui.adsabs.harvard.edu/abs/2020MNRAS.494.3177M} {494, 3177}

\bibitem[\protect\citeauthoryear{{Mondal}, {Raychaudhuri}, {Dewangan}  \&
  {Beri}}{{Mondal} et~al.}{2022}]{Mondal2022MNRAS.516.1256M}
{Mondal} A.~S.,  {Raychaudhuri} B.,  {Dewangan} G.~C.,   {Beri} A.,  2022,
  \mn@doi [\mnras] {10.1093/mnras/stac2321}, \href
  {https://ui.adsabs.harvard.edu/abs/2022MNRAS.516.1256M} {516, 1256}

\bibitem[\protect\citeauthoryear{Mukherjee, Bult, vanderKlis  \&
  Bhattacharya}{Mukherjee et~al.}{2015}]{Mukherjee10.1093/mnras/stv1542}
Mukherjee D.,  Bult P.,  vanderKlis M.,   Bhattacharya D.,  2015, \mn@doi
  [Monthly Notices of the Royal Astronomical Society] {10.1093/mnras/stv1542},
  452, 3994

\bibitem[\protect\citeauthoryear{{Muno}, {Remillard}  \& {Chakrabarty}}{{Muno}
  et~al.}{2002}]{Muno2002ApJ...568L..35M}
{Muno} M.~P.,  {Remillard} R.~A.,   {Chakrabarty} D.,  2002, \mn@doi [\apjl]
  {10.1086/340269}, \href
  {https://ui.adsabs.harvard.edu/abs/2002ApJ...568L..35M} {568, L35}

\bibitem[\protect\citeauthoryear{{O'Neill}, {Sood}, {Kuulkers}  \& {van der
  Klis}}{{O'Neill} et~al.}{2001}]{2001ASPC..251..396O}
{O'Neill} P.~M.,  {Sood} R.~K.,  {Kuulkers} E.,   {van der Klis} M.,  2001, in
  {Inoue} H.,  {Kunieda} H.,  eds,  Astronomical Society of the Pacific
  Conference Series Vol. 251, New Century of X-ray Astronomy. p.~396

\bibitem[\protect\citeauthoryear{{Oosterbroek}, {Barret}, {Guainazzi}  \&
  {Ford}}{{Oosterbroek} et~al.}{2001}]{Oosterbroek2001A&A...366..138O}
{Oosterbroek} T.,  {Barret} D.,  {Guainazzi} M.,   {Ford} E.~C.,  2001, \mn@doi
  [\aap] {10.1051/0004-6361:20000028}, \href
  {https://ui.adsabs.harvard.edu/abs/2001A&A...366..138O} {366, 138}

\bibitem[\protect\citeauthoryear{{Pan}, {Zhang}, {Song}, {Wang}, {Li}  \&
  {Yang}}{{Pan} et~al.}{2018}]{Pan2018MNRAS.480..692P}
{Pan} Y.~Y.,  {Zhang} C.~M.,  {Song} L.~M.,  {Wang} N.,  {Li} D.,   {Yang}
  Y.~Y.,  2018, \mn@doi [\mnras] {10.1093/mnras/sty1851}, \href
  {https://ui.adsabs.harvard.edu/abs/2018MNRAS.480..692P} {480, 692}

\bibitem[\protect\citeauthoryear{{Peirano} \& {M{\'e}ndez}}{{Peirano} \&
  {M{\'e}ndez}}{2022}]{Peirano2022MNRAS.513.2804P}
{Peirano} V.,  {M{\'e}ndez} M.,  2022, \mn@doi [\mnras]
  {10.1093/mnras/stac1071}, \href
  {https://ui.adsabs.harvard.edu/abs/2022MNRAS.513.2804P} {513, 2804}

\bibitem[\protect\citeauthoryear{{Ponti} et~al.,}{{Ponti}
  et~al.}{2015}]{Ponti2015MNRAS.446.1536P}
{Ponti} G.,  et~al., 2015, \mn@doi [\mnras] {10.1093/mnras/stu1853}, \href
  {https://ui.adsabs.harvard.edu/abs/2015MNRAS.446.1536P} {446, 1536}

\bibitem[\protect\citeauthoryear{{Popham} \& {Sunyaev}}{{Popham} \&
  {Sunyaev}}{2001}]{Popham2001ApJ...547..355P}
{Popham} R.,  {Sunyaev} R.,  2001, \mn@doi [\apj] {10.1086/318336}, \href
  {https://ui.adsabs.harvard.edu/abs/2001ApJ...547..355P} {547, 355}

\bibitem[\protect\citeauthoryear{{Pringle} \& {Rees}}{{Pringle} \&
  {Rees}}{1972}]{Pringle1972A&A....21....1P}
{Pringle} J.~E.,  {Rees} M.~J.,  1972, \aap, \href
  {https://ui.adsabs.harvard.edu/abs/1972A&A....21....1P} {21, 1}

\bibitem[\protect\citeauthoryear{{Rea}, {Stella}, {Israel}, {Matt}, {Zane},
  {Segreto}, {Oosterbroek}  \& {Orlandini}}{{Rea}
  et~al.}{2005}]{Rae2005MNRAS.364.1229R}
{Rea} N.,  {Stella} L.,  {Israel} G.~L.,  {Matt} G.,  {Zane} S.,  {Segreto} A.,
   {Oosterbroek} T.,   {Orlandini} M.,  2005, \mn@doi [\mnras]
  {10.1111/j.1365-2966.2005.09646.x}, \href
  {https://ui.adsabs.harvard.edu/abs/2005MNRAS.364.1229R} {364, 1229}

\bibitem[\protect\citeauthoryear{{Revnivtsev} \& {Gilfanov}}{{Revnivtsev} \&
  {Gilfanov}}{2006}]{Revnivtsev2006A&A...453..253R}
{Revnivtsev} M.~G.,  {Gilfanov} M.~R.,  2006, \mn@doi [\aap]
  {10.1051/0004-6361:20053964}, \href
  {https://ui.adsabs.harvard.edu/abs/2006A&A...453..253R} {453, 253}

\bibitem[\protect\citeauthoryear{Risaliti et~al.,}{Risaliti
  et~al.}{2013}]{Risaliti}
Risaliti G.,  et~al., 2013, \mn@doi [Nature] {10.1038/nature11938}, 494,
  449–451

\bibitem[\protect\citeauthoryear{{Sanna} et~al.,}{{Sanna}
  et~al.}{2010}]{Sanna2010MNRAS.408..622S}
{Sanna} A.,  et~al., 2010, \mn@doi [\mnras] {10.1111/j.1365-2966.2010.17145.x},
  \href {https://ui.adsabs.harvard.edu/abs/2010MNRAS.408..622S} {408, 622}

\bibitem[\protect\citeauthoryear{{Schulz}, {Hasinger}  \& {Truemper}}{{Schulz}
  et~al.}{1989}]{Schulz1989A&A...225...48S}
{Schulz} N.~S.,  {Hasinger} G.,   {Truemper} J.,  1989, \aap, \href
  {https://ui.adsabs.harvard.edu/abs/1989A&A...225...48S} {225, 48}

\bibitem[\protect\citeauthoryear{{Shakura} \& {Sunyaev}}{{Shakura} \&
  {Sunyaev}}{1973}]{Shakura1973A&A....24..337S}
{Shakura} N.~I.,  {Sunyaev} R.~A.,  1973, \aap, \href
  {https://ui.adsabs.harvard.edu/abs/1973A&A....24..337S} {24, 337}

\bibitem[\protect\citeauthoryear{{Sharma}, {Sharma}, {Jain}  \&
  {Dutta}}{{Sharma} et~al.}{2020}]{Sharma2020MNRAS.496..197S}
{Sharma} P.,  {Sharma} R.,  {Jain} C.,   {Dutta} A.,  2020, \mn@doi [\mnras]
  {10.1093/mnras/staa1516}, \href
  {https://ui.adsabs.harvard.edu/abs/2020MNRAS.496..197S} {496, 197}

\bibitem[\protect\citeauthoryear{{Shirey}, {Bradt}  \& {Levine}}{{Shirey}
  et~al.}{1999}]{Shirey1999ApJ...517..472S}
{Shirey} R.~E.,  {Bradt} H.~V.,   {Levine} A.~M.,  1999, \mn@doi [\apj]
  {10.1086/307188}, \href
  {https://ui.adsabs.harvard.edu/abs/1999ApJ...517..472S} {517, 472}

\bibitem[\protect\citeauthoryear{{Sibgatullin} \& {Sunyaev}}{{Sibgatullin} \&
  {Sunyaev}}{2000}]{Sibgatullin2000AstL...26..699S}
{Sibgatullin} N.~R.,  {Sunyaev} R.~A.,  2000, \mn@doi [Astronomy Letters]
  {10.1134/1.1323277}, \href
  {https://ui.adsabs.harvard.edu/abs/2000AstL...26..699S} {26, 699}

\bibitem[\protect\citeauthoryear{{Sleator} et~al.,}{{Sleator}
  et~al.}{2016}]{Sleator2016ApJ...827..134S}
{Sleator} C.~C.,  et~al., 2016, \mn@doi [\apj] {10.3847/0004-637X/827/2/134},
  \href {https://ui.adsabs.harvard.edu/abs/2016ApJ...827..134S} {827, 134}

\bibitem[\protect\citeauthoryear{{Tauris} \& {van den Heuvel}}{{Tauris} \& {van
  den Heuvel}}{2006}]{tauris_vandenheuvel_2006}
{Tauris} T.~M.,  {van den Heuvel} E.~P.~J.,  2006, Formation and evolution of
  compact stellar X-ray sources.
Cambridge University Press, p. 623–666, \mn@doi{10.1017/CBO9780511536281.017}

\bibitem[\protect\citeauthoryear{{Tomsick} et~al.,}{{Tomsick}
  et~al.}{2018}]{Tomsick2018ApJ...855....3T}
{Tomsick} J.~A.,  et~al., 2018, \mn@doi [\apj] {10.3847/1538-4357/aaaab1},
  \href {https://ui.adsabs.harvard.edu/abs/2018ApJ...855....3T} {855, 3}

\bibitem[\protect\citeauthoryear{{Ueda}, {Murakami}, {Yamaoka}, {Dotani}  \&
  {Ebisawa}}{{Ueda} et~al.}{2004}]{Ueda2004ApJ...609..325U}
{Ueda} Y.,  {Murakami} H.,  {Yamaoka} K.,  {Dotani} T.,   {Ebisawa} K.,  2004,
  \mn@doi [\apj] {10.1086/420973}, \href
  {https://ui.adsabs.harvard.edu/abs/2004ApJ...609..325U} {609, 325}

\bibitem[\protect\citeauthoryear{{Ueda}, {Mitsuda}, {Murakami}  \&
  {Matsushita}}{{Ueda} et~al.}{2005}]{2005ApJ...620..274U}
{Ueda} Y.,  {Mitsuda} K.,  {Murakami} H.,   {Matsushita} K.,  2005, \mn@doi
  [\apj] {10.1086/426933}, \href
  {https://ui.adsabs.harvard.edu/abs/2005ApJ...620..274U} {620, 274}

\bibitem[\protect\citeauthoryear{{Verner}, {Ferland}, {Korista}  \&
  {Yakovlev}}{{Verner} et~al.}{1996}]{1996ApJ...465..487V}
{Verner} D.~A.,  {Ferland} G.~J.,  {Korista} K.~T.,   {Yakovlev} D.~G.,  1996,
  \mn@doi [\apj] {10.1086/177435}, \href
  {https://ui.adsabs.harvard.edu/abs/1996ApJ...465..487V} {465, 487}

\bibitem[\protect\citeauthoryear{{White}, {Lightman}  \& {Zdziarski}}{{White}
  et~al.}{1988}]{1988ApJ...331..939W}
{White} T.~R.,  {Lightman} A.~P.,   {Zdziarski} A.~A.,  1988, \mn@doi [\apj]
  {10.1086/166611}, \href
  {https://ui.adsabs.harvard.edu/abs/1988ApJ...331..939W} {331, 939}

\bibitem[\protect\citeauthoryear{{Wijnands}, {M{\'e}ndez}, {van der Klis},
  {Psaltis}, {Kuulkers}  \& {Lamb}}{{Wijnands}
  et~al.}{1998}]{1998ApJ...504L..35W}
{Wijnands} R.,  {M{\'e}ndez} M.,  {van der Klis} M.,  {Psaltis} D.,  {Kuulkers}
  E.,   {Lamb} F.~K.,  1998, \mn@doi [\apjl] {10.1086/311564}, \href
  {https://ui.adsabs.harvard.edu/abs/1998ApJ...504L..35W} {504, L35}

\bibitem[\protect\citeauthoryear{{Wilms}, {Allen}  \& {McCray}}{{Wilms}
  et~al.}{2000}]{2000ApJ...542..914W}
{Wilms} J.,  {Allen} A.,   {McCray} R.,  2000, \mn@doi [\apj] {10.1086/317016},
  \href {https://ui.adsabs.harvard.edu/abs/2000ApJ...542..914W} {542, 914}

\bibitem[\protect\citeauthoryear{{Zdziarski}, {Szanecki}, {Poutanen},
  {Gierli{\'n}ski}  \& {Biernacki}}{{Zdziarski}
  et~al.}{2020}]{Zdziarski2020MNRAS.492.5234Z}
{Zdziarski} A.~A.,  {Szanecki} M.,  {Poutanen} J.,  {Gierli{\'n}ski} M.,
  {Biernacki} P.,  2020, \mn@doi [\mnras] {10.1093/mnras/staa159}, \href
  {https://ui.adsabs.harvard.edu/abs/2020MNRAS.492.5234Z} {492, 5234}

\bibitem[\protect\citeauthoryear{Zhang}{Zhang}{2007}]{Zhang_10.1007/978-3-540-74713-0_114}
Zhang C.,  2007, in Aschenbach B.,  Burwitz V.,  Hasinger G.,   Leibundgut B.,
  eds, Relativistic Astrophysics Legacy and Cosmology -- Einstein's. Springer
  Berlin Heidelberg, Berlin, Heidelberg, pp 490--492

\bibitem[\protect\citeauthoryear{{Zhang} \& {Kojima}}{{Zhang} \&
  {Kojima}}{2006}]{Zhang2006MNRAS.366..137Z}
{Zhang} C.~M.,  {Kojima} Y.,  2006, \mn@doi [\mnras]
  {10.1111/j.1365-2966.2005.09802.x}, \href
  {https://ui.adsabs.harvard.edu/abs/2006MNRAS.366..137Z} {366, 137}

\bibitem[\protect\citeauthoryear{{Zhu}, {L{\"u}}, {Wang}  \& {Wang}}{{Zhu}
  et~al.}{2012}]{Zhu2012PASP..124..195Z}
{Zhu} C.,  {L{\"u}} G.,  {Wang} Z.,   {Wang} N.,  2012, \mn@doi [\pasp]
  {10.1086/664833}, \href
  {https://ui.adsabs.harvard.edu/abs/2012PASP..124..195Z} {124, 195}

\bibitem[\protect\citeauthoryear{{van der Klis}}{{van der
  Klis}}{2000}]{2000ARA&A..38..717V}
{van der Klis} M.,  2000, \mn@doi [\araa] {10.1146/annurev.astro.38.1.717},
  \href {https://ui.adsabs.harvard.edu/abs/2000ARA&A..38..717V} {38, 717}

\makeatother
\end{thebibliography}







\bsp	
\label{lastpage}
\end{document}